\documentclass[aps, prfluids, showkeys, showpacs, notitlepage, amsmath, amssymb, floatfix, onecolumn, longbibliography, tightenlines, 12pt]{revtex4-2}

\usepackage{amsmath} 
\usepackage{amssymb}
\usepackage{hyperref}
\usepackage[normalem]{ulem}
\usepackage{mathtools} 
\usepackage{mathptmx}
\usepackage{dcolumn, booktabs}
\newcolumntype{d}[1]{D{.}{.}{#1}}
\newcommand\mc[1]{\multicolumn{1}{c}{#1}}

\newcommand{\Smp}[2]{S_{{#1}}^{#2\text{pt}}}   
\newcommand{\Sm}[1]{S_{{#1}}^{\text{ }}\,}     
\newcommand{\lse}[2]{\xi_{{#1}}^{#2\text{pt}}} 

\begin{document}

\title{Multi-time structure functions and the Lagrangian scaling of turbulence}
\author{Sof\'{\i}a Angriman}
\email[Corresponding author: ]{sangriman@df.uba.ar}
\author{Pablo D. Mininni}
\author{Pablo J. Cobelli}
\affiliation{Universidad de Buenos Aires, Facultad de Ciencias Exactas y
Naturales, Departamento de F\'\i sica, \& IFIBA, CONICET, Ciudad Universitaria,
Buenos Aires 1428, Argentina}

\begin{abstract}
We define and characterize multi-time Lagrangian structure functions using data stemming from two swirling flows with mean flow and turbulent fluctuations: A Taylor-Green numerical flow, and a von K\'arm\'an laboratory experiment. Data is obtained from numerical integration of tracers in the former case, and from three-dimensional particle tracking velocimetry measurements in the latter. Multi-time statistics are shown to decrease the contamination of large scales in the inertial range scaling. A time scale at which contamination from the mean flow becomes dominant is identified, with this scale separating two different Lagrangian scaling ranges. The results from the multi-time structure functions also indicate that Lagrangian intermittency is not a result of large-scale flow effects. The multi-time Lagrangian structure functions can be used without prior knowledge of the forcing mechanisms or boundary conditions, allowing their application in different flow geometries.
\end{abstract}

\maketitle 

\section{Introduction}
Structure functions, the statistical moments of velocity increments as a function of the spatial or time scale, are a fundamental tool to study fully developed turbulent flows and to assess how strong the intermittency is, i.e., how much the statistics deviate from the scaling predicted by Kolmogorov's 1941 (K41) theory \cite{She_1994}. In the context of K41 theory, for very large Reynolds numbers and at scales (both spatial and temporal) that are small compared to the typical energy injection scales, the statistics of the flow should be self-similar and universal, and therefore independent of the forcing mechanisms.  However, deviations from this prediction in the form of intermittency have been observed even at very high Reynolds numbers in homogeneous and isotropic turbulent flows, indicating that its occurrence is not a finite Reynolds effect \cite{Iyer_2020}. Such deviations were also found in fractional, low order exponents of structure functions, which are less affected by extreme events and by departures from universality \cite{Chen_2005}. In order to quantify the deviations from self-similarity, the proper identification of the bounds of the inertial range in structure functions is key, and in the Eulerian frame the third-order moment of the longitudinal velocity increments has been recognized as a less ambiguous indicator of inertial-range scaling \cite{Anselmet_1984}.

In the last decades the Lagrangian approach to turbulence, being the most natural for problems of mixing and transport, has received more attention, as both numerical and experimental methods became available to track individual particles at high rates and in large volumes \cite{Toschi_2009, Voth_2017}. In spite of this, the Lagrangian characterization of turbulence and the understanding of intermittency in the Lagrangian frame remain challenging. Tracers' velocity structure functions are known to display short ranges compatible with Kolmogorov scaling, with an amplitude and width that grows slowly with the Reynolds number \cite{Mordant_2001, Biferale_2008, Sawford_2011, Falkovich_2012}. Furthermore, large-scale flows have been found to affect both small-scale Lagrangian and Eulerian turbulent statistics \cite{Mordant_2002, Ouellette_2006, Blum_2010, Blum_2011, Angriman_2020}. More recently, long Lagrangian trajectories were shown to retain information of the large-scale flow topology \cite{Angriman_2021}. Evaluating how these effects impact on the observed scaling is important for the development of accurate models of Lagrangian turbulence \cite{Beck_2007, Viggiano_2020}. As an example, statistical theories derived directly from the Navier-Stokes equations without fitting parameters result in measurable relations for the exponents of Lagrangian velocity structure functions, which are expected to hold even in the absence of scaling \cite{Zybin_2008}. It has been speculated that Lagrangian intermittency may result from the coupling of slow and fast time scales, arising from correlations between acceleration and velocity increments \cite{Wilczek_2013}. And it was also reported that the squared acceleration of tracers, coarse-grained over viscous time scales, recovers Gaussian statistics \cite{Bentkamp_2019}.

Many tools have been developed to improve the observed scaling and to reduce mean-flow and anisotropy effects induced by forcing mechanisms. In the Eulerian frame, extended self-similarity (ESS) \cite{Benzi_1993} has been widely used to analyze scaling properties at finite Reynolds numbers. Angle averaging of structure functions has also been proposed as a way to recover isotropy in Eulerian statistics of arbitrarily forced flows \cite{Eyink_2002, Taylor_2003}. Indeed, performing an SO(3) decomposition to velocity correlation or structure functions has been used to disentangle anisotropic from isotropic contributions to the scaling in numerical simulations \cite{Taylor_2003, Biferale_2005, Martin_2010, Iyer_2017}. In the Lagrangian frame, considering data conditioned to small spatial regions is also a usual practice to recover more homogeneous and isotropic statistics \cite{Blum_2010, Volk_2011}, and also to reduce sweeping effects from the large-scale flow on the particles' velocity statistics \cite{Angriman_2020}. The Lagrangian acceleration spectrum has been used as a better proxy for inertial-range scaling, as it is less affected by the mean flow and finite Reynolds number effects \cite{Sawford_2011}. Complementarily, the Hilbert-Huang transform methodology has been generalized and applied to single-particle velocity data to extract the hierarchy of the Lagrangian scaling exponents without resorting to structure functions or the ESS procedure \cite{Huang_2013}.

Higher-degree velocity differences, for instance using wavelet analysis \cite{Chevillard_2012} and coarse-graining \cite{Faller_2021}, have been proposed as unbiased quantities to study Eulerian and Lagrangian statistics, and as a way to detect high-order singularities in the velocity field. In the Eulerian frame, they provide a robust and straightforward technique to quantify intermittency in a signal with a steep power spectrum, such as those observed in surface wave turbulence \cite{Falcon_2010}. A related tool, the so-called multi-point Eulerian structure functions, become useful as well to filter out the contribution of smooth and large-scale fields as in the case of studies of sub-ion-scale fluctuations in plasma turbulence, where the turbulent scaling is sub-leading and the power spectra are also steep \cite{Cerri_2019, Wang_2020}.  The importance of multi-point measurements has also been pointed out as a general way to distinguish between turbulent fluctuations and waves in space plasmas \cite{Matthaeus_2019}.

In this work we construct and characterize multi-time differences (or high-degree differences) of the Lagrangian velocity and study their moments, employing particle data from two turbulent swirling flows: Taylor-Green direct numerical simulations, and a von K\'arm\'an laboratory experiment.  In both systems sweeping is observed to affect the particles' statistics, as the tracers are affected by the large-scale counter-rotating eddies that make up the mean, macroscopic flow. At this point, it is worth noting that these flows display a mean flow (or circulation) with a complex geometry, together with large- and small-scale fluctuations. This contrasts with the paradigmatic case of homogeneous and isotropic turbulence (HIT). Distinguishing between mean circulation and large-scale fluctuations is not trivial, specially in the Lagrangian framework and for flows in which the mean flows are not stationary and average to zero in volume \cite{Ravelet_2004, Mininni_2014}. How to disentangle these contributions in realistic flows is a long-standing problem in turbulence (see, e.g., a method proposed by Batchelor \cite{Batchelor_1957} and a recent application \cite{Viggiano_2021}). Our main motivation is then to reduce the contribution resulting from large-scale flow components, which in our flows entangle mean circulation and random sweeping (the case of HIT, without a mean flow, will be also considered), and to evaluate Lagrangian intermittency once this effect has been alleviated. When compared to their two-times counterparts, we observe that multi-time Lagrangian structure functions display a significant reduction of the large-scale flow contribution on the tracers' statistics. This allows us to identify a time scale at which contamination from the mean flow becomes dominant, and which separates two different Lagrangian scaling ranges. The results from the multi-time analysis also suggest that Lagrangian intermittency is not a result of large-scale flow effects. The method presented here is suitable for other types of flow geometries, as {\it a priori} knowledge of forcing mechanisms or boundary conditions is not required for its application. 

\section{Multi-time Lagrangian structure functions}

In this Section we introduce the multi-time technique, following the presentation for the multi-point Eulerian structure functions given in \citet{Cho_2019}. The goal is to remove slow (or large-scale) variations in a signal, in order to isolate the turbulent-like fluctuations.

For the sake of consistency with our Lagrangian case (and without loss of generality) we take that signal to be a time series of the particle velocity, denoted herein by $v(t)$. This signal is considered to be composed of short-scale (i.e., fast) fluctuations $v_S(t)$ with correlation time $\tau_S$, and large scale (i.e., slow) fluctuations $v_L(t)$, so that $v(t) = v_L(t) + v_S(t)$. Firstly, the fast fluctuations are assumed to have turbulent-like statistics, in the sense that its power spectral density follows some power law scaling, with the largest power present at time scale $\tau_S$, and with less power present as the time scale decreases. Secondly, the mean of $v_S(t)$ is taken to be equal to $0$ when averaging over time intervals larger that $\tau_S$, i.e., $\langle v_S(t) \rangle _{\Delta t > \tau_S} = 0$. Central differences using multiple times (or ``points'' in the time series) are then defined as 
\begin{equation}
\Delta v^{(2\text{pt})}_\tau (t) = v(t+\tau) - v(t),
\end{equation}
\begin{equation}
\Delta v^{(3\text{pt})}_\tau (t) = v(t+\tau) - 2v(t) + v(t - \tau),
\end{equation}
\begin{equation}
\Delta v^{(4\text{pt})}_\tau (t) = v(t-\tau) - 3v(t) + 3v(t+\tau) - v(t + 2\tau),
\end{equation}
and so on. Given a number of times $n$, the corresponding structure function of order $p$ is then
\begin{equation}
\Smp{p}{n}(\tau) = \Lambda^{(n\text{pt})} \langle |\Delta v^{(n\text{pt})}_\tau|^p\rangle,
\label{eq:Smp}
\end{equation}
where $\langle\cdot\rangle$ indicates an average over the time $t$, and $\Lambda^{(2\text{pt})} = 1$, $\Lambda^{(3\text{pt})} = 1/3$, and $\Lambda^{(4\text{pt})} = 1/10$ (see Appendix A of Ref.~\cite{Cho_2019} for a general expression of $\Delta v_\tau^{(n\text{pt})}$ and $\Smp{p}{n}$). Strictly speaking, the velocity differences defined above are high-degree differences, which are always centered around the same time instant $t$. Since we are interested in studying high-order moments of said differences, and for consistency with the name used in plasma physics in the Eulerian frame of ``multi-point structure" functions, we refer to the structure functions defined in Eq.~(\ref{eq:Smp}) as multi-time structure functions.

\begin{figure}[b]
  {\includegraphics[width=0.85\textwidth]{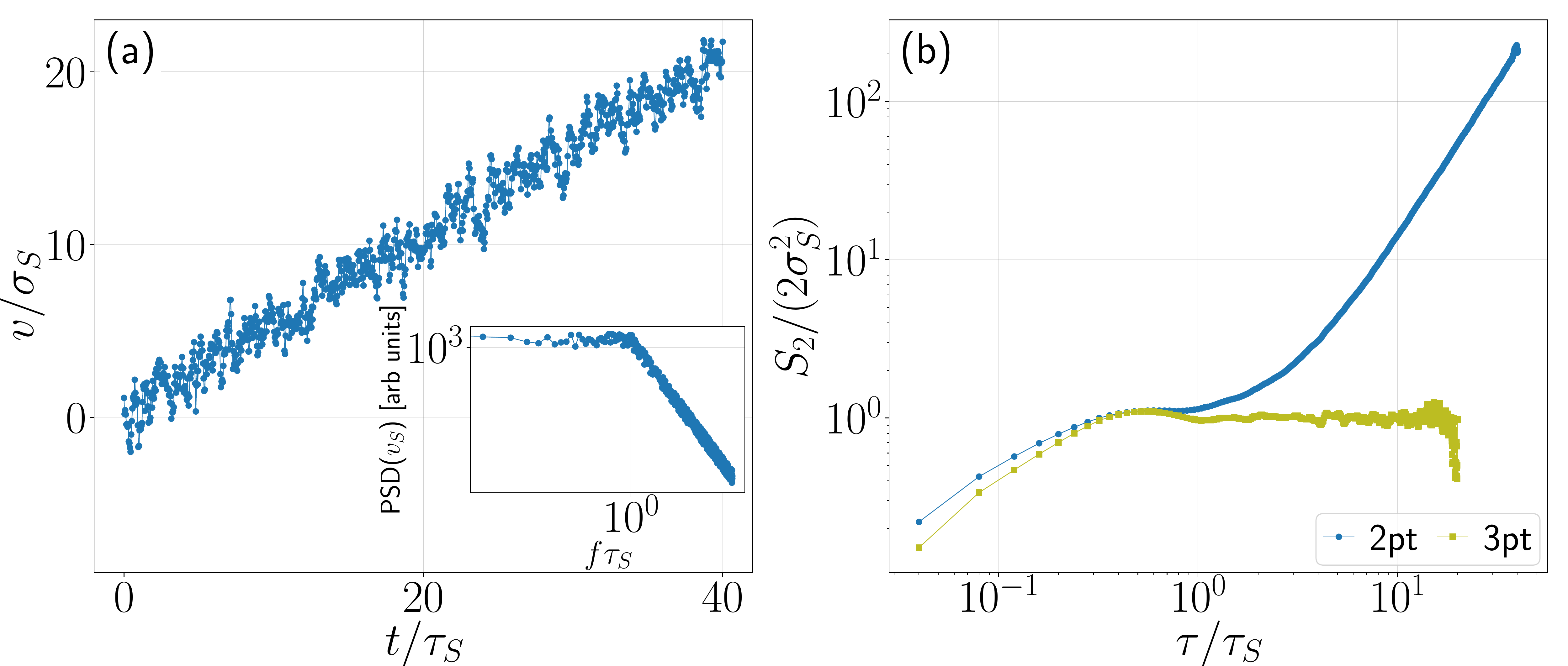}}%
  \hfill
  \caption{Illustration of the multi-time method: (a) A synthetically generated signal $v$, composed of fast (small-scale) fluctuations $v_S$ superimposed on a slow (large-scale) trend $v_L$ (chosen in the example as a first degree polynomial), and normalized by the standard deviation of $v_S$, denoted by $\sigma_S$. The inset shows the power spectral density of $v_S$. (b) Normalized second-order structure function using 2 and 3 times.}
  \label{fig:synth_MP}
\end{figure}

Let's now consider a random signal with a given and fixed variance, and with a slow trend. If for $\tau \gtrsim \tau_S$ the difference associated to the slow (large-scale) variations is smaller than the one associated to the fast (small-scale) variations, i.e., $|\Delta v_{\tau,L}^{(n\text{pt})}| < |\Delta v_{\tau,S}^{(n\text{pt})}|$, then the structure function using $n$ times is able to substantially remove the contributions corresponding to the slow modulation. In particular, it can be shown that a structure function using $n$ times (or points) can completely remove slow polynomial contributions of order $n-2$ \cite{Cho_2019}. Focusing on the case $p=2$, this implies that while structure functions using a small number of times will still show contamination from the slow variation, for sufficiently large $n$ and for $\tau \gtrsim \tau_S$ a plateau with amplitude proportional to the squared standard deviation of the small scale fluctuations $\sigma_S^2$ should become visible for our random signal. To illustrate this behavior we consider a synthetically generated random signal superimposed on a slow variation (in this example, a 1st degree polynomial), as shown in Fig.~\ref{fig:synth_MP}(a). The inset shows the power spectral density of the fast fluctuations $v_S(t)$ in log-log scale, which displays a power-law decay for $f\tau_S \geq 1$. In Fig.~\ref{fig:synth_MP}(b), the second-order structure function calculated using 2 and 3 times are shown, compensated by $2 \sigma_S^2$. Note that the two-times second-order structure function, $\Smp{2}{2}$, presents a plateau only at a very reduced range of time lags. In contrast, the three-times counterpart, $\Smp{2}{3}$, successfully removes the effect of the slow variations, displaying a significantly broader plateau which extends beyond a decade and whose amplitude is proportional to $\sigma_S^2$.

\section{Description of the datasets}

While multi-point structure functions (or related techniques using, e.g., wavelets) were used with Eulerian data in plasma physics and wave turbulence \cite{Falcon_2010, Cho_2019}, with a few exceptions  they have not been used to analyze Lagrangian statistics \cite{Chevillard_2012}, and higher-order moments of the statistics were not considered. We thus consider two different datasets (albeit with similar large-scale flows in spite of differences in the boundary conditions and the forcing mechanism \cite{Mininni_2014, Angriman_2020}) to study whether multi-time methods can reduce the effect of the large-scale flow in Lagrangian statistics, and recover previously observed scaling laws. The two datasets stem from two sources: a von K\'arm\'an (VK) laboratory experiment, and Taylor-Green (TG) direct numerical simulations (DNSs).

The VK experiment consists of two facing disks of diameter $D = 19$~cm, each fitted with $8$ straight blades, separated by a distance of $20$~cm and contained in a square cross-section cell with side of $20$~cm. As the propellers counter rotate, they agitate the fluid, which in this case is water. This generates two large counter-rotating circulation cells, which produce (on average) a strong shear layer in the mid-plane of the propellers, as well as a secondary circulation in the axial direction. In this way we obtain a three-dimensional (3D) fully developed turbulent flow in a confined region of space of dimensions $(20\times 20\times 20)~\text{cm}^3$. The flow has a mean, macroscopic structure which is anisotropic: the large scale structures in the horizontal direction (i.e., parallel to disks' plane) are larger that the structures in the axial direction. For this study we consider two values for the rotation frequency of the disks $f_0^\text{VK}$, namely $0.83$~Hz and $1.25~$Hz (or equivalently, $50$~rpm and $75$~rpm, respectively). These choices were made to have Reynolds numbers in the experiment of the same order as those obtained in the DNSs described below. The flow is seeded with tracer particles, which are neutrally buoyant (density $1$~g~cm$^{-3}$) polyethylene microspheres of diameter $d=250-300~\mu$m (Cospheric). The measurement of their dynamics is done using Particle Tracking Velocimetry (PTV), employing two high-speed cameras at sampling frequency $f_S^\text{VK} \approx 2/\tau_\eta$, with $\tau_\eta = (\nu / \varepsilon)^{1/2}$ being the Kolmogorov time scale. The 3D individual evolution of each particle is tracked in a region of size $(16\times16\times16)~\text{cm}^3$ around the geometrical center of the cell, whose size is comparable to the entire experimental volume. It comprises both the shear-dominated region of the flow and the areas closer to the propellers, where energy is injected. After several realizations of the experiment there are $\mathcal{O}(10^4)$ trajectories with a mean duration of $0.34/f_0^\text{VK}$. From the individual trajectories, the instantaneous velocity is then computed after applying a Gaussian filter. For further details on the experimental setup and the measurement technique see
Ref.~\cite{Angriman_2020} (see also \cite{Mordant_2004, Ouellette_2006, Volk_2011} for studies on other VK experiments).

For the TG simulations, we performed DNSs of the
incompressible Navier-Stokes equations, where the turbulence is sustained by an external volumetric forcing $\mathbf{F}$. The equations are solved in a 3D cubic, $2\pi$-periodic domain using a parallel pseudospectral method with the GHOST code \cite{Mininni_2011, Rosenberg_2020}, with spatial resolutions of $512^3$, $768^3$, and $1024^3$ grid points. The mechanical forcing $\mathbf{F}$ is based
on the TG flow \cite{Brachet_1983}, with Cartesian components
\begin{equation}
    F_x = F_0~\text{sin}(k_F x)~\text{cos}(k_F y)~\text{cos}(k_F z),\quad
    F_y = -F_0~\text{cos}(k_F x)~\text{sin}(k_F y)~\text{cos}(k_F z),\quad
    F_z = 0,
\end{equation}
and with forcing wave number $k_F = 1$. This forcing consists of a periodic array of counter-rotating large-scale vortices. Note that it is anisotropic: it is similar in the horizontal ($x$, $y$) direction but it injects no energy into the $z$ component of the velocity.  As a result of the symmetries that the flow presents in a statistical sense, while still being anisotropic, the full domain can be split into eight cells of size $\pi^3$. In each of these cells, the resulting flow strongly resembles the large-scale geometry of the VK flow: two counter-rotating large-scale vortices separated by a shear layer in the midplane; these similarities with the experiment hold for both the Eulerian and Lagrangian descriptions, and have been exploited in several studies (see \cite{Mininni_2014, Angriman_2020} and references therein). In the turbulent steady state of this flow, $10^6$ Lagrangian particles are evolved in time along with the flow according to
\begin{equation}
    \frac{d\mathbf{x}_P}{dt} = \mathbf{u}(\mathbf{x}_P,t),
\end{equation}
where $\mathbf{x}_P$ is the position of the Lagrangian tracer at time $t$, and $\mathbf{u}(\mathbf{x}_P,t)$ is the velocity of the fluid element at position $\mathbf{x}_P(t)$. The instantaneous position, velocity and acceleration of each particle are computed as it evolves. Note these particles are ideal tracers, as they do not interact with each other nor do they affect the flow's dynamics.

In order to characterize both flows, and considering their anisotropy, we use the one-component r.m.s.~tracers' velocity $U$, associated to motions in the horizontal plane. That is, we define $U$ as
\begin{equation}
    U = \sqrt{\langle v_x^2 + v_y^2\rangle/2 },
\end{equation}
where $v_i$ is the tracers' velocity component along the $i$ direction, and the brackets $\langle\cdot\rangle$ denote averages over time and trajectories (in previous studies we observed that this way of estimating the r.m.s.~velocity in the TG and VK flow results in better comparisons between non-dimensional parameters in DNSs and experiments \cite{Angriman_2020}; taking into account the three components of the velocity does not change significantly the estimation of $U$,  resulting in values $4$ to $8\%$ smaller).
We also define the integral length scale $L_0$ based on the scale where energy is injected into the systems. In the experiments, $L_0^\text{VK} = D = 0.19$~m, the diameter of the disks. In the TG DNSs, the squared norm of the perpendicular components of the forcing wave vector (with respect to $\hat{z}$) satisfy $\mathbf{k_{F\perp}} \cdot \mathbf{k_{F\perp}} = 2$, so that $L_0^{\text{TG}} = 2\pi/|\mathbf{k_{F\perp}}| = 2\pi/\sqrt{2}$ (we use $\mathbf{k_{F\perp}}$ for consistency with the experiment, where we only consider the diameter of the disks as the energy injection scale, ignoring the vertical flow dependence).  In this way, an integral time can then be defined as
\begin{equation}
    T_0 = \frac{L_0}{U}.
\end{equation}

The energy injection rate $\varepsilon$ is directly available in the DNSs. In the VK flow we estimate it as
\begin{equation}
    \varepsilon = \alpha \frac{U^3}{L_0},
\label{eq:eps}
\end{equation} 
where $\alpha$ is a non-dimensional proportionality constant. We use the TG DNSs to estimate $\alpha$ by computing $\varepsilon/(U^3/L_0)$, and observe that it varies by less than $5\%$ in the range of Reynolds numbers explored. Thus, we use $\alpha=1.69$, which corresponds to the TG DNS value using $1024^3$ grid points, i.e., for the simulation with the largest Reynolds number available. 

\begin{table}[t!]
\caption{Values of the parameters for both laboratory experiments and numerical simulations. DNSs values are dimensionless. The label corresponds to the frequency of the propellers in the experiments (in rpm), and to the spatial resolution in the DNSs. $U$ denotes the root-mean-square value of the particles' horizontal velocity. $T_0$ stands for the integral time, and $\tau_L$ represents the particle correlation times (in units of $T_0$). The kinematic viscosity is symbolized by $\nu$, and $\varepsilon$ is the energy injection rate.
$\textrm{Re}$ is an integral-scale Reynolds number based on the particles' velocity $U$, 
and $\textrm{Re}_\lambda = \sqrt{15 U^4/(\nu \varepsilon)}$ is the Taylor microscale Reynolds number.}
\label{table1}
\begin{tabular*}{\textwidth}{l @{\extracolsep{\fill}} r c c c c c c c c c}
\hline \hline
\hfill
Flow                  & Label    & $U$    & $T_0$ & $\tau_L/T_0$ & $\nu$               & $\varepsilon$       &  $\textrm{Re}$  & $\textrm{Re}_\lambda$  \\
                      &          & [m/s]  & [s]   &              & $\times 10^{-6}$ [ m$^2$/s] & $\times 10^{-2}$ [ m$^2$/s$^3$] & & \\
\hline
\addlinespace[1mm]
VK                    & 50~rpm   & $0.11$ & $1.8$ & $0.38$       & $1$                 & $1.2$                   & $2.1\times10^{4}$  & $430$         \\
VK                    & 75~rpm   & $0.17$ & $1.1$ & $0.38$       & $1$                 & $4.3$                   & $3.2\times10^{4}$  & $530$         \\
\addlinespace[1mm]
\hline
\addlinespace[1mm]
TG                    & $512^3$  & $0.84$ & $5.3$ & $0.40$       & $675$               & $24$                    & $5.5\times10^{3}$  & $215$         \\
TG                    & $768^3$  & $0.85$ & $5.2$ & $0.42$       & $450$               & $24$                    & $8.4\times10^{3}$  & $270$         \\
TG                    & $1024^3$ & $0.85$ & $5.2$ & $0.44$       & $300$               & $23$                    & $1.3\times10^{4}$  & $335$         \\
\addlinespace[0.5mm]
\hline \hline
\end{tabular*}
\end{table}

We can then estimate two Reynolds numbers based on Lagrangian quantities, in order to compare experiments and DNSs on an equal footing. An
integral Reynolds number based on the tracers' velocity $U$ is defined as
\begin{equation}
    \textrm{Re} = \frac{U L_0}{\nu},
\end{equation}
where $\nu$ is the kinematic viscosity. This number should not be confused with a Reynolds number of the flow at the particle size, as we only consider tracers in this study. Also, using the definition of $\varepsilon$ in Eq.~(\ref{eq:eps}), the Taylor-based Reynolds number can be estimated as
\begin{equation}
    \textrm{Re}_\lambda = \sqrt{\frac{15~U^4}{\nu~\varepsilon}}.
\end{equation}

Finally, besides the integral time $T_0$ we can also estimate a Lagrangian correlation time directly from the tracers' dynamics. The particles' velocity correlation function $R_L^{(i)}(\tau)= C_v(\tau)/C_v(0) = \langle v_i(t)~v_i(t+\tau)\rangle/\langle v_i^2(t)\rangle$ is computed for each of the Cartesian components of the tracers' velocities. The one-dimensional Lagrangian correlation time $\tau_L^{(i)}$ is then estimated from its corresponding correlation function as the time of the first zero-crossing. Then, for each dataset, the correlation time is defined as $\tau_L = (\tau_L^{(x)} + \tau_L^{(y)})/2$. Note that we base this estimation on the correlation function of the horizontal components of the velocity, as the largest structures in the flow lie in the $x-y$ plane, and as a result the correlation time of the tracers is dominated by these large-scale eddies. In the experiments, the correlation function is also affected by the different lengths of the trajectories: the shorter the curves, the faster the decay observed in $R_L(\tau)$. In particular, in the experiments where particles can get in or out of the measurement volume, the trajectories do not have the same extension in time, and the distribution of lengths of trajectories has an exponential decay with the trajectory length. To take into account the effect of this on the estimation of the correlation time, we sub-sampled the numerical data to obtain a similar distribution of trajectories lengths. To this end we randomly picked a subset of trajectories from the $10^6$ trajectories available in each run, and each one was cropped from an initial time chosen randomly, up to a final time, so that the lengths of all cropped trajectories followed the same distribution of lengths observed in the experiments. From this synthetically sub-sampled data we computed the velocity correlation function and the time lag of the first zero crossing, and compared its value with $\tau_L$, the zero crossing of $R_L(\tau)$ using the complete dataset with the long trajectories. This allowed us to estimate the finite-time (or finite sampling) correction to the time $\tau_L$ measured in the experiments.

All the resulting parameters for the two experimental and the three
numerical datasets are listed in table~\ref{table1}. Unless stated otherwise, the results presented in the following sections correspond to the largest Reynolds numbers available in each setup: the VK experiment at $75$~rpm, and the TG DNS with $1024^3$ grid points.

\section{Results}

\subsection{Multi-time Lagrangian second-order velocity structure functions \label{sec:second_ord}}

\begin{figure}[t!]
  {\includegraphics[width=1\textwidth]{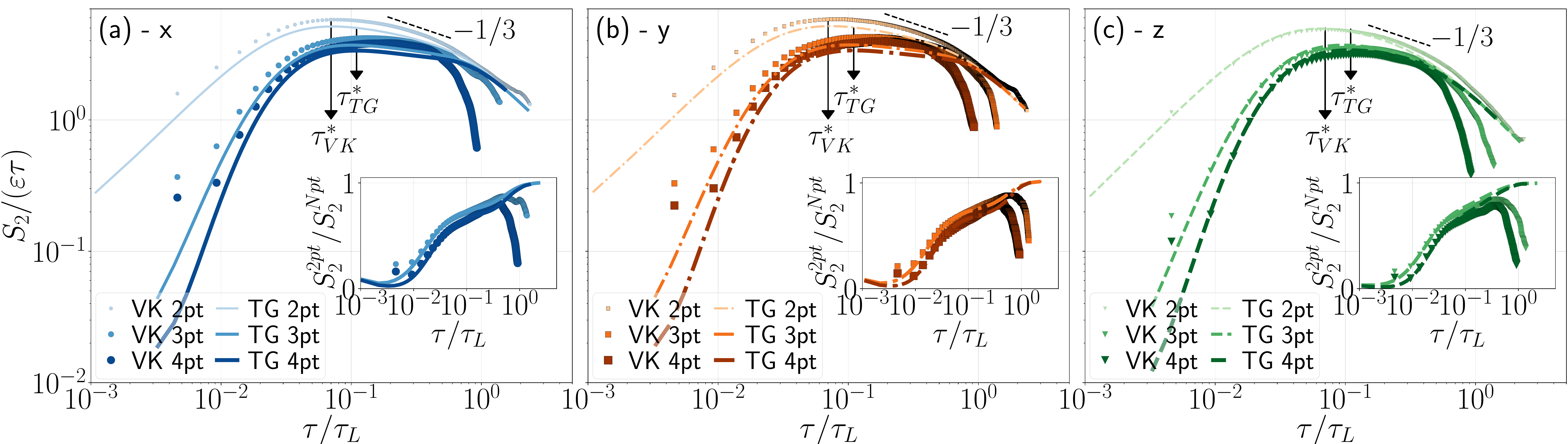}}
  \caption{Second-order tracers' velocity structure functions for the VK experiment (dots)  and for the TG DNSs (solid curves) using 2, 3 and 4 times; for each of the Cartesian components $x$, $y$, and $z$ in panels (a), (b), and (c) respectively. A power law with exponent $-1/3$ -which corresponds to sweeping by the mean flow given by $S_2 \propto \tau^{2/3}$- is indicated by the dashed line. The time scale $\tau^*$ where this contribution is comparable to inertial range scaling is indicated by arrows (normalized by $\tau_L$) for both datasets. The insets show $\Smp{2}{2}/\Smp{2}{3}$ and $\Smp{2}{2}/\Smp{2}{4}$ for each Cartesian component. Labels are the same as in the main panels.}
  \label{fig:S2_tau}
\end{figure}

We begin by computing the multi-time second-order velocity structure functions $S_2(\tau)$, using 2, 3, and 4 times. Fig.~\ref{fig:S2_tau} shows these curves, compensated by the inertial range prediction $S_2(\tau) \sim \varepsilon \tau$, for each of the Cartesian components of the velocity, $x$, $y$, and $z$ (panels a, b, and c respectively) in the VK experiment and the TG simulations. As the number of times employed in the calculation of the structure function increases from 2 to 3, a scaling compatible with $\tau$ appears more clearly. Not only the plateau is broader, but also its limits are steeper, allowing for a better and less ambiguous identification of the range of time scales consistent with inertial-range behavior. By further increasing the number of times used from 3 to 4, one can identify more clearly the limits of the region for which the scaling $\Sm{2}(\tau) \propto \tau$ is approximately satisfied. Also, and as observed in \cite{Angriman_2020} for 2-times Lagrangian structure functions, and as will be shown below for higher-order structure functions, the Lagrangian statistical properties of the VK experiment and the TG flow are similar.

In a previous study we showed that the effect of sweeping by the large-scale flow (present both in the VK experiment and the TG flow) is partially the reason for the poor scaling of $\Smp{2}{2}$ with the time lag \cite{Angriman_2020}. Indeed, we have observed that in these flows $\Smp{2}{2}(\tau)$ presents a scaling proportional to $\tau^{2/3}$ for time lags of the order of $\tau_L$, which is consistent with sweeping by large-scale eddies. A power law with this exponent (which results in $-1/3$ when compensated by $\tau$) is shown as a reference in Fig.~\ref{fig:S2_tau}. In \cite{Angriman_2020} this was also verified by conditioning the statistics to different spatial regions where the large-scale flow has larger or smaller intensity (with the associated improvement in the inertial-range scaling in the latter case). Note that random sweeping cannot be neglected even for the idealized case of HIT \cite{Chen_1989}, and its effect cannot be removed simply by moving to a Lagrangian frame \cite{Kraichnan_1965b}. The contamination of the Lagrangian scaling by sweeping caused by the mean flow is further supported here by the fact that when we consider multi-time structure functions, such as $\Smp{2}{3}$ or $\Smp{2}{4}$, which are expected to remove slow modulations, they display a scaling proportional to $\tau$ at time lags $\tau \lesssim \tau_L$, while $\Smp{2}{2}$ does not.

In the Eulerian description of turbulence, it is usual to relate the second-order structure function at a scale $\ell$ with the energy contained in eddies with characteristic size $r < \ell$. However, eddies larger than $\ell$ make a non-negligible contribution to $S_2$ of order $\ell^2 (\partial_r u|_{r=\ell})^2$ (see, e.g., \cite{Davidson_2015}). In the Lagrangian framework, this argument can be recast as follows: For a sufficiently smooth velocity field, $S_2 (\tau) = \Smp{2}{2}(\tau) = \langle (\Delta v_\tau)^2 \rangle = \langle [v(t+\tau) - v(t)]^2 \rangle$ can be approximated as
\begin{equation}
    S_2(\tau) \sim \tau^2
    \left< \left( \frac{\partial v}{\partial t} \right)^2 \right>_{T>\tau} ,
\end{equation}
where the subindex indicates the average must be computed over times $T$ longer than $\tau$. In the case of a rough field with Kolmogorov scaling, in contrast, the eddies with a turnover time $T \leq \tau$ make a contribution to $S_2(\tau)$ which is proportional to $\tau$, and then
\begin{equation}
    S_2(\tau) \sim \tau + \tau^2 \left< \left( \frac{\partial v}{\partial t} \right)^2 \right>_{T>\tau}.
\end{equation}
As a result, in this expression the term proportional to $\tau$ is associated to the contribution from fluctuations with timescale $\tau$ or faster, while the term proportional to $\tau^2$ comes from the eddies with turnover time slower than $\tau$ (i.e., from the smooth part of the field). The latter term can be readily estimated as
\begin{equation}
\tau^2 \left< \left(\frac{\partial v}{\partial \tau}\right)^2 \right>_{T>\tau} \sim \tau^2 \frac{\delta U_\tau^2}{\tau^2}, 
\end{equation}
where $\delta U_\tau$ is the typical variation of the slowly evolving component of the flow over times $\tau$. As a Lagrangian particle moves through the fluid over a time $\tau$ in the inertial range, the smooth, slow (and large-scale) flow should not change significantly as its turnover time is larger than $\tau$. Consequently, the variation $\delta U_\tau$ seen by the fluid element must result from the displacement $\ell$ of the particle across the large-scale flow structure, with $\ell \sim U \tau$. It follows that $\delta U_\tau \sim \delta U_\ell\vert_{\ell=U \tau} \sim \ell^{1/3}\vert_{U \tau} \sim (U \tau)^{1/3}$, assuming Kolmogorov scaling. Finally, including quantities missing using dimensional analysis, as well as the missing prefactors, we expect that
\begin{equation}
    S_2(\tau) \approx C_0 \varepsilon \tau + \beta \varepsilon^{2/3} U^{2/3} \tau^{2/3},
\end{equation}
where $C_0$ is the Lagrangian constant and $\beta \approx 2 $ is the Kolmogorov's constant \cite{Ishihara_2009}. Note that sweeping of the particles by a large-scale flow results in a $\propto \tau^{2/3}$ scaling of the 2-times Lagrangian second-order structure function. This expression is compatible with the behavior of $\Smp{2}{2}(\tau)$ in Fig.~\ref{fig:S2_tau}: for long time increments $\Smp{2}{2}(\tau) \propto \tau^{2/3}$, and only in a very short interval $\Smp{2}{2}(\tau) \propto \tau$. The effect of sweeping by the smooth, large-scale flow becomes of the same order as that of the rough scales for a time lag $\tau^* \sim (\beta/C_0)^3 U^2/\varepsilon$. It is worth noting that, for a given flow, the range of time scales dominated by sweeping is approximately independent of the Reynolds number.

Using $\beta = 2$ and estimating $C_0$ as the maximum value of
$\Smp{2}{2}/(\varepsilon\tau)$, the estimated values of $\tau^*/\tau_L$ for the VK and the TG data are $\tau^*/\tau_L \approx 7\times 10^{-2}$ and $\approx 1.1\times 10^{-1}$, respectively. These estimations are compatible with the time lags for which $\Smp{2}{3}$ and $\Smp{2}{4}$ present the most significant differences with $\Smp{2}{2}$ (i.e., while the plateau in $\Smp{2}{2}$ compensated by $\varepsilon \tau$ is only visible around that time lag, the same plateau is visible for a much wider range of time lags in the compensated versions of both $\Smp{2}{3}$ and $\Smp{2}{4}$). In other words, for $\tau \gtrsim \tau^*$ the range of time lags for which large-scale sweeping becomes $\mathcal{O}(1)$ is clearly visible in the 2-times second-order structure function, and is not present in the 3- and 4-times structure functions. This indicates that indeed the multi-times structure functions are able to reduce the contamination in the inertial range scaling by the large-scale flow.

\begin{figure}[b!]
  {\includegraphics[width=1\textwidth,trim={0 2.35cm 0 0},clip]{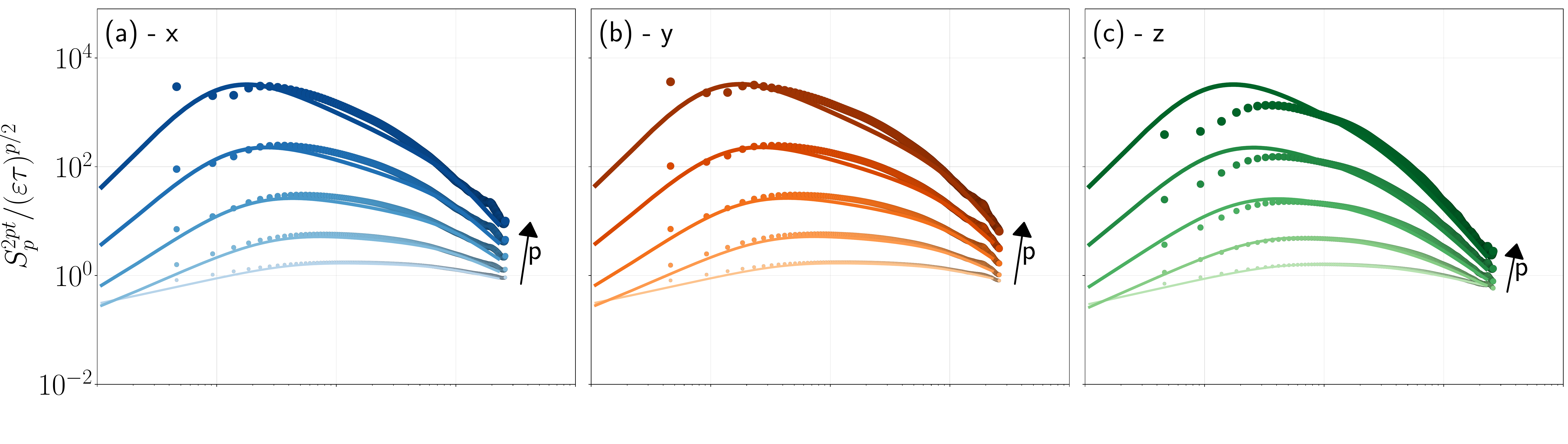}}%
  \hfill
  {\includegraphics[width=1\textwidth,trim={0 2.35cm 0 0},clip]{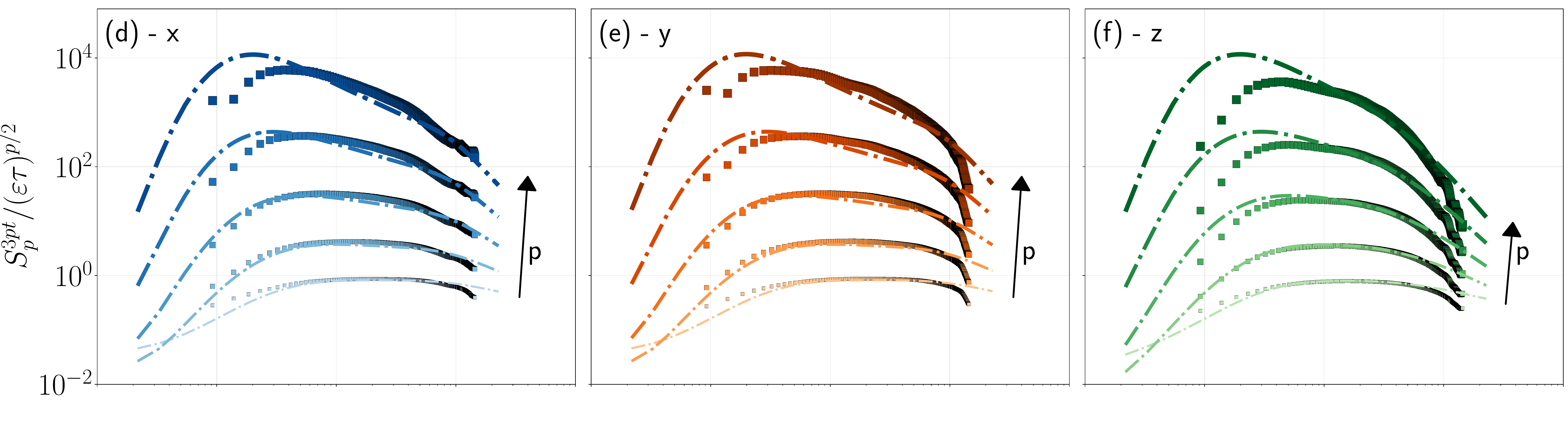}}%
  \hfill
   {\includegraphics[width=1\textwidth]{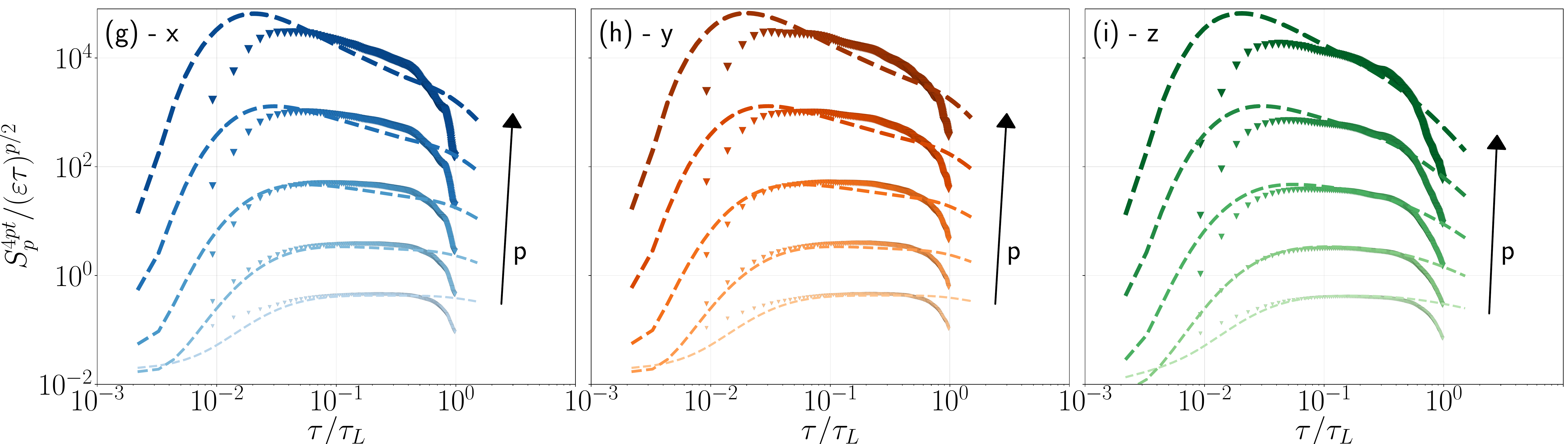}}%
	\caption{Multi-time Lagrangian velocity structure functions $\Smp{p}{n}$ of orders $p=1$ to $5$, compensated by the non-intermittent Lagrangian inertial-range prediction $S_p(\tau) \propto \tau^{p/2}$, for tracers in the TG and VK datasets (represented by lines and individual markers, respectively). The top, middle and bottom row panels correspond to calculations using 2-, 3-, and 4-times, respectively. Each column gathers one individual Cartesian component of the particles' velocity. In all panels, the arrows indicate increasing order $p$.}
  \label{fig:Sp_mp}
\end{figure}

The insets in each panel of Fig.~\ref{fig:S2_tau} show $\Smp{2}{2}/\Smp{2}{3}$ and $\Smp{2}{2}/\Smp{2}{4}$. Notably, the relative change in the amplitude of the structure functions as the number of times used is increased is similar in the VK experiment and in the TG DNS. In all cases there is a considerable amplitude decrease with $n$, in the dissipative range as well as near the integral time scale. In particular, for time lags comparable with $\tau_L$, while the numerical curves collapse for 3- and 4-times, the experimental curves display a sharp decrease in amplitude. This difference may be related to how localized the forcing is in Fourier space in each of the two setups considered, which also affects the degree of contamination by the mean flow at intermediate scales. Furthermore, in the experiment the sharp decrease in amplitude near $\tau_L$ is also related to finite-volume effects, as the computation of structure functions for large $\tau$ requires long trajectories. These correspond to particles that are slow enough to remain in the field of view for a sufficiently long period of time, which introduces a bias towards smaller velocity variances. The increase in the number of times used to compute the multi-time structure functions requires even longer curves, which further increases the drop in the amplitude. Note also the difference between VK and TG near $\tau_L$ in this respect: in the simulation, where the total number of trajectories is the same for all time lags and where the sampling is uniform, no finite sampling effects are present and the use of 3 or 4 times only shortens the total range of time-lags available.

Finally, it is worth pointing out that even in HIT, which does not present a mean circulation, inertial range scaling is also affected by the large-scale velocity fluctuations. Appendix \ref{app:HIT} presents the multi-time second-order structure function using 2, 3 and 4 times in a $1024^3$ simulation of HIT with random forcing. A cleaner Kolmogorov-like scaling for all of the Cartesian components of the Lagrangian velocity is observed (similar results are obtained for higher-order moments). This further supports the idea that the presented multi-time structure functions are able to reduce large-scale flow effects, irrespective of their nature (i.e., mean global circulation effects, or random sweeping by large-scale velocity fluctuations).

\subsection{Higher-order Lagrangian structure functions}

As higher-order moments of the velocity differences are considered, identifying a scaling that may differ or may be compatible with the non-intermittent inertial-range prediction, $\Sm{p} \propto \tau^{p/2}$, becomes harder at finite Reynolds numbers. In light of our previous results for $p=2$, it is natural to wonder whether using multi-time statistics also improves the scaling for higher values of $p$. Figure~\ref{fig:Sp_mp} shows the 2-, 3-, and 4-times structure functions of order $p$, compensated by the inertial-range prediction, for both experiments and simulations. These results are arranged as follows. Each row of panels corresponds to a different number of times $n$ used in the computation: 2 at the top, 3 at the center, and 4 at the bottom. Columns, in turn, gather each Cartesian component of the tracers' velocity from which the corresponding structure functions are calculated: $x$ on the left, $y$ on the center, and $z$ on the right.  Within each individual panel, the multi-time structure functions $\Smp{p}{n}$ of orders $p = 1$ to $5$ are displayed in ascending order from bottom to top.

In the first row in Fig.~\ref{fig:Sp_mp}, corresponding to the standard $\Smp{p}{2}$ structure functions, very narrow plateaus (or, otherwise, power-law ranges compatible with inertial-range scaling corrected by intermittency) are visible in all-order compensated structure functions. Also, note that as $p$ is increased, the range of time lags compatible with inertial-range scaling seems to shift towards smaller time lags $\tau$. This apparent shift of the inertial range towards shorter (or smaller) scales, also observed in Eulerian structure functions, was previously reported in \cite{Xu_2006}. In spite of the short ranges compatible with power law scaling, note that both the VK and TG datasets display similar behavior of all the structure functions, as also expected from previous comparisons between the two flows \cite{Angriman_2020}.

By comparing curves of the same order $p$ between panels in different rows in Fig.~\ref{fig:Sp_mp} (i.e., with increasing number of times $n$, from top to bottom), it becomes evident that increasing the number of times improves the power-law scaling of the structure functions. Remarkably, this effect can be seen in both the VK experiments and the TG DNSs. Also, the apparent shift of the range of scales compatible with inertial-range scaling becomes either negligible or very small. Focusing on the case with $p=2$, and as already discussed in Sec.~\ref{sec:second_ord}, increasing $n$ from 2 to 3 or 4 results in a wider compensated inertial-range plateau. And when values of $p \neq 2$ are considered, ranges compatible with power-law behavior can be easily identified in the same inertial range of $\Smp{2}{n}$ (for $n=3$ and 4), albeit without perfect $\sim \tau^{p/2}$ scaling. As will be discussed in the next section, this is the result of intermittency corrections to the scaling of higher-order structure functions. By comparing the curves with $n=3$ and $n=4$, it is also clear that increasing the number of times improves the scaling specially for large values of $p$. 

Finally, we can compare the different columns in Fig.~\ref{fig:Sp_mp}. While the structure functions of the $x$ and $y$ components of the velocity are similar (for the same values of $n$ and $p$), the structure functions of the $z$-velocity component show some differences (specially for $n=2$). This is the result of the flow anisotropy in both the VK and TG flows: the large-scale $z$ velocity does not feature the same amplitude as the horizontal components, as the forcing mechanism is not isotropic. Also note that the curves in Fig.~\ref{fig:Sp_mp}(c) display the largest differences between the DNS and the experiment. As it was shown in \cite{Angriman_2020}, the level of large-scale anisotropy in the VK and TG flows is not quite the same, and this has an impact in $\Smp{2}{2}(\tau)$, specially in the $z$ direction.

\subsection{Local scaling exponents and Lagrangian intermittency}

\begin{figure}[t!]
  {\includegraphics[width=1\textwidth,trim={0 2.35cm 0 0},clip]{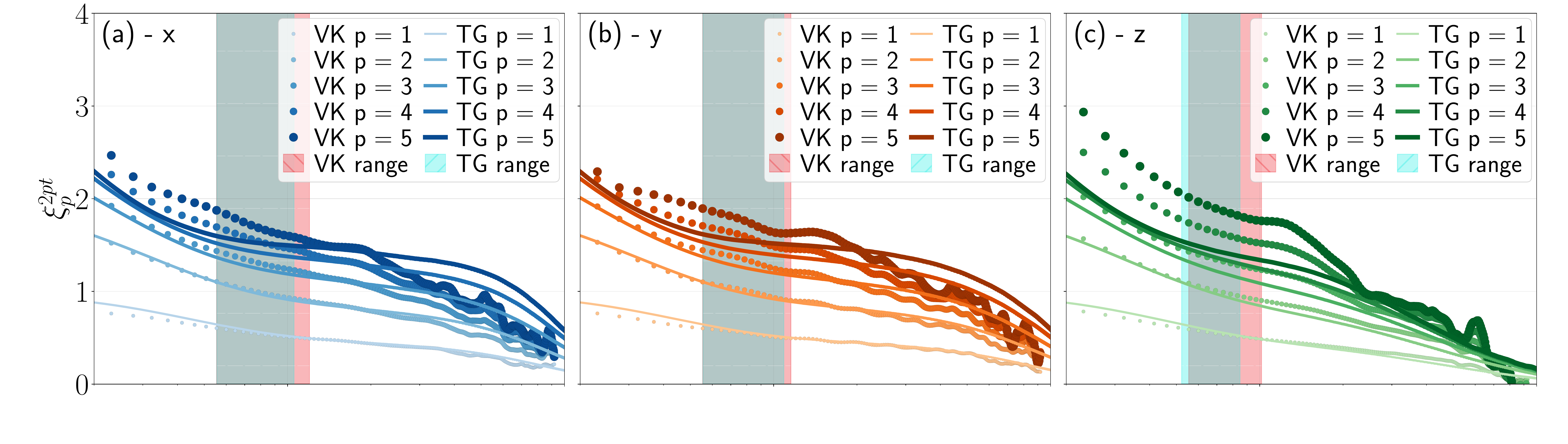}}%
  \hfill
  {\includegraphics[width=1\textwidth,trim={0 2.35cm 0 0},clip]{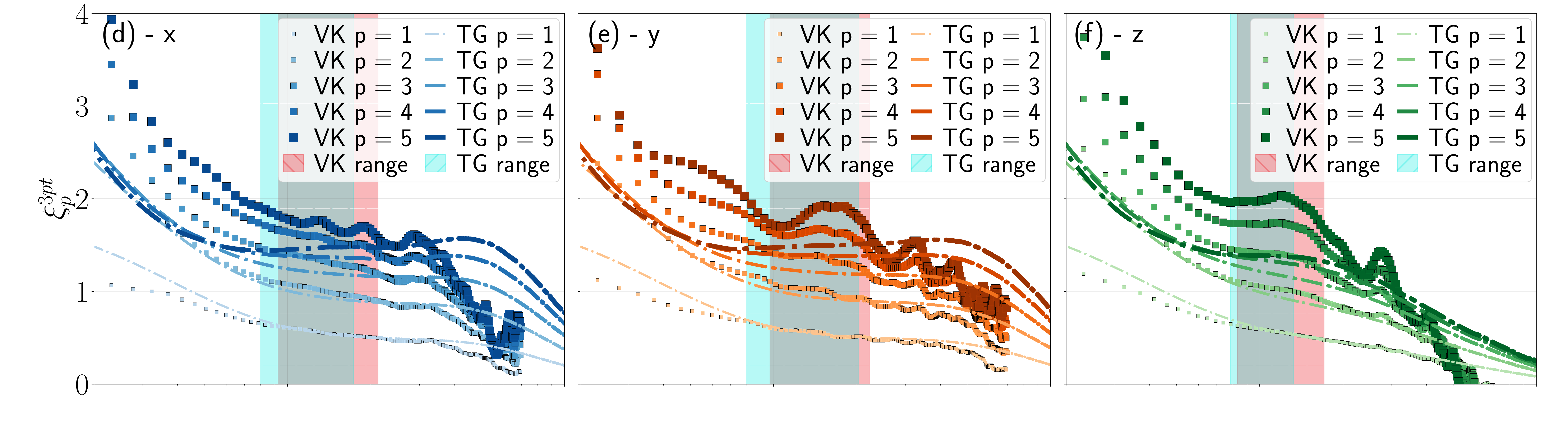}}%
  \hfill
  {\includegraphics[width=1\textwidth]{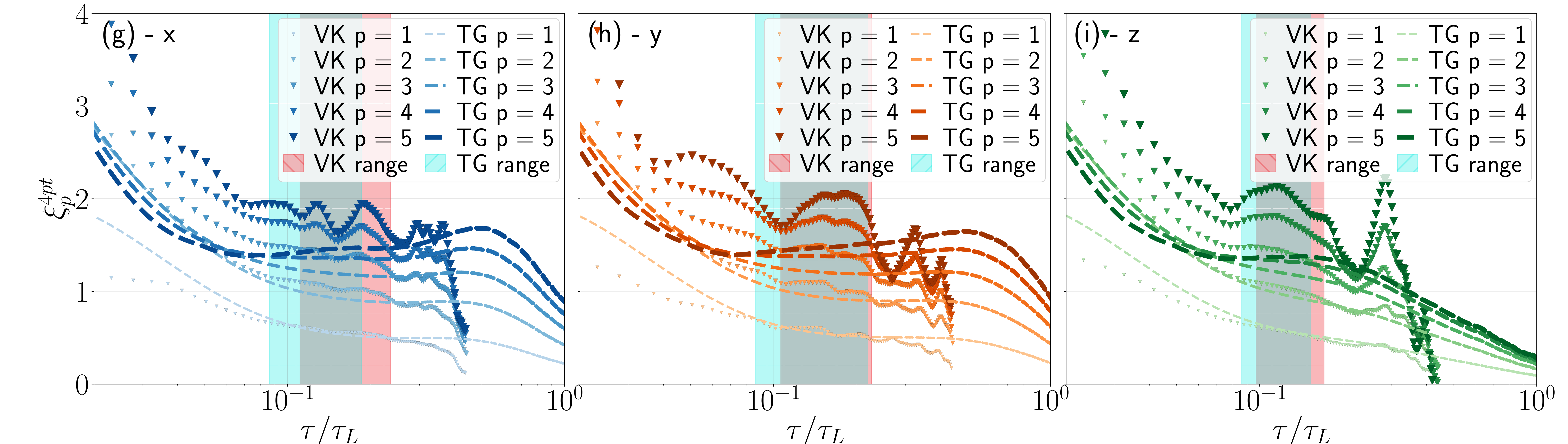}}%
  \hfill
  \caption{Logarithmic derivative of the multi-time structure functions of orders $p=1$ to 5; for tracers in the VK and TG datasets  (represented by markers and lines, respectively). Panels at the top, middle, and bottom rows correspond to $\xi_p^\text{npt}$ calculated using 2, 3, and 4 times, respectively. Each column shows one individual Cartesian component of the particles' velocity ($x$, $y$, and $z$, respectively). The shaded areas correspond to the inertial range, for practical purposes identified as the range in $\tau/\tau_L$ for which $0.9 \leq \lse{2}{n} \leq 1.1$.}
  \label{fig:LSE_mp}
\end{figure}

We now evaluate the local scaling exponents (LSE) of order $p$, $\lse{p}{n}$, via the logarithmic derivative of the corresponding $n$-time structure function
\begin{equation}
	\lse{p}{n}(\tau) = \frac{\text{d~log}[\Smp{p}{n}(\tau)]}{\text{d~log}(\tau)},
\label{eq:LSE}
\end{equation}
for $p=1$ to $5$, and $n = 2$, 3, and 4. In particular, we are interested in the
values of the LSEs within the Lagrangian inertial range. For practical purposes, to identify the inertial range we will use the range of time scales for which $S_2(\tau) \sim \varepsilon \tau$. More precisely, we identify it as the range $[\tau_1, \tau_2]$, where $\tau_{1,2}$ are such that $\lse{2}{n}(\tau_2) = 0.9$ and $\lse{2}{n}(\tau_1) = 1.1$, i.e., they bound symmetrically the range where the second-order scaling exponent presents a value compatible with Kolmogorov scaling ($\xi_2 = 1$). We verified that using more stringent conditions on $\lse{2}{n}(\tau_1)$ and $\lse{2}{n}(\tau_2)$ did not change the determination of the other exponents within the error bars. The LSEs thus determined are shown in the different panels of Fig.~\ref{fig:LSE_mp}. The presentation of these results is similar to that in Fig.~\ref{fig:Sp_mp}, with the three columns corresponding to estimations for the $x$, $y$, and $z$ velocity components, and the three rows corresponding to increasing number of times used in the computation of the structure functions. In each panel, the shaded areas indicate the location of the Lagrangian inertial range, identified as previously described.

\begin{figure}[b!]
  {\includegraphics[width=1\textwidth]{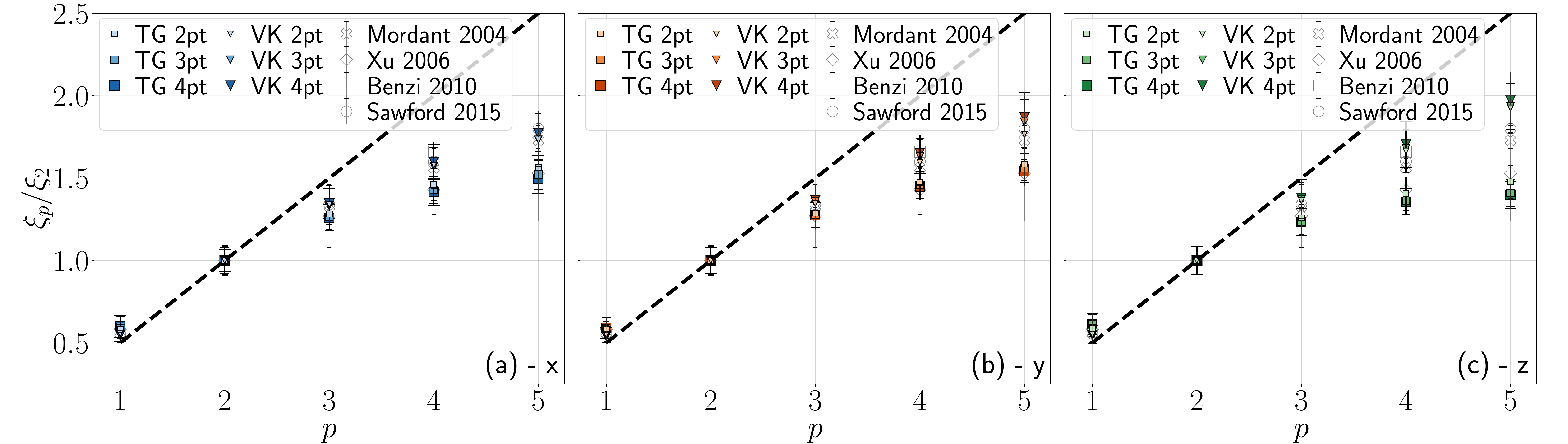}}
  \hfill
  \caption{Inertial-range local scaling exponents of order $p$, normalized by the exponent corresponding to $p=2$, and computed from multi-time structure functions for the TG and VK data. The exponents are compared with other results from experiments and simulations with different flows from Refs.~\cite{Mordant_2004, Xu_2006, Benzi_2010, Sawford_2015}. From left to right: (a) $\xi_p/\xi_2$ for the $x$ component of the velocity, (b) for $y$, and (c) for $z$. In all panels the dashed line indicates the non-intermittent (i.e., self-similar) Kolmogorov scaling.}
  \label{fig:exp_mp}
\end{figure}

In Fig.~\ref{fig:exp_mp} the inertial range exponents (normalized by the second-order exponent) are shown, for each number of times $n$ considered in the computation of the structure functions. These exponents, and their uncertainty, are computed by taking the mean, and the standard deviation, respectively, of $\xi_p$ in the range $[\tau_1, \tau_2]$. The dashed line in all panels corresponds to $p/2$, the expected non-intermittent (i.e., self-similar) Lagrangian inertial-range scaling. The exponents for the VK and TG data are also compared with results by \citet{Mordant_2004, Xu_2006, Benzi_2010}, and \citet{Sawford_2015}, all computed using extended self-similarity for different turbulent flows in laboratory experiments or in numerical simulations. Our exponents are in close agreement with the exponents reported in the literature. The multi-time structure functions allow for a better estimation of the scaling exponents at finite Reynolds numbers, as they result in a broader range of time scales with power-law scaling, specially for high-order (in $p$) structure functions. The scaling exponents of the multi-time structure functions computed using ESS (listed in appendix \ref{app:ESS}, with their uncertainties) also yield values which are consistent with those previously reported in the literature, and present a similar behavior to the LSEs computed using Eq.~(\ref{eq:LSE}).

Note that in Fig.~\ref{fig:exp_mp} no significant differences in the deviation of the scaling exponents from the self-similar 
prediction can be identified between the exponents computed using 2-, 3-, and 4-times statistics. In other words, the level of Lagrangian intermittency is the same in all cases. It has been argued that the origin of Lagrangian intermittency may be associated to the coupling of the fast eddies with the mean flow or with long-time correlations in the dynamics, and in particular, with very slow modulations with time scales longer than the Lagrangian correlation time (see, e.g., \cite{Mordant_2004, Wilczek_2013}). As the 3- and 4-times structure functions were shown to significantly reduce the contamination by the mean flow, while still maintain the same level of intermittency, this result suggests that intermittency is more of an intrinsic feature of inertial-range Lagrangian turbulence. 

\begin{figure}[t!]
  {\includegraphics[width=0.5\textwidth]{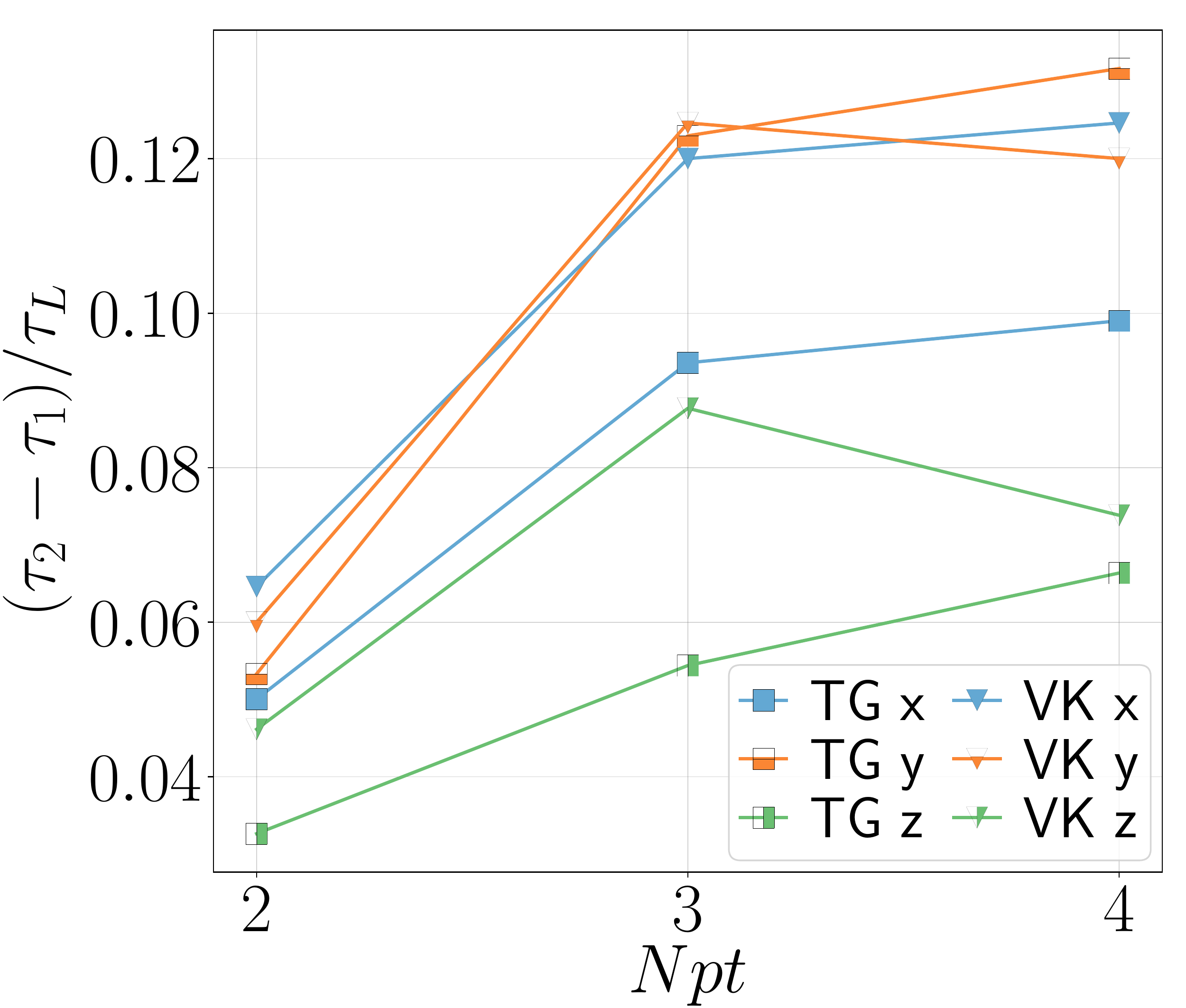}}%
  \hfill
  \caption{Absolute width of the inertial range, normalized by the Lagrangian correlation time, as a function of the number of times $n$ used to compute the structure functions, for each Cartesian component of the velocity and for the VK experiment (triangles) and the TG DNS (squares).}
  \label{fig:width_mp}
\end{figure}

\begin{table*}[b!]
\caption{Local scaling exponents (LSEs) $\xi_p$ of orders $p=1,\ldots,5$ for the $x$, $y$, and $z$ components of the Lagrangian velocity in the TG and VK datasets, computed from logarithmic derivatives of structure functions using 2, 3 and, 4 times. Uncertainties are listed below each LSE value.}
\label{table2}
\renewcommand*{\arraystretch}{1.4}		
\newcommand*{\cpm}{{\scriptstyle\pm}}	
\scriptsize
\begin{tabular*}{\textwidth}{@{\extracolsep{\fill}} c *{9}{d{1.3}} *{9}{d{1.3}} }
\hline \hline \hfill
		  &           &            &            &           & \mc{TG}    &            &            &            &            &           &            &           &           & \mc{VK}    &           &           &            &           \\
\cmidrule[0.4pt]{2-10} \cmidrule[0.4pt]{11-19} 
		  &           & \mc{2pt} &            &           & \mc{3pt} &            &            & \mc{4pt} &            &           & \mc{2pt} &           &           & \mc{3pt} &           &           & \mc{4pt} &           \\
\cmidrule[0.4pt]{2-4} \cmidrule[0.4pt]{5-7} \cmidrule[0.4pt]{8-10} \cmidrule[0.4pt]{11-13} \cmidrule[0.4pt]{14-16} \cmidrule[0.4pt]{17-19}
\mc{$p$}  & \mc{$x$}  & \mc{$y$}   & \mc{$z$}   & \mc{$x$}  & \mc{$y$}   & \mc{$z$}   & \mc{$x$}   & \mc{$y$}   & \mc{$z$}   & \mc{$x$}  & \mc{$y$}   & \mc{$z$}  & \mc{$x$}  & \mc{$y$}   & \mc{$z$}  & \mc{$x$}  & \mc{$y$}   & \mc{$z$}  \\
\cmidrule[0.4pt]{1-19}
1         & 0.57      & 0.56       & 0.58       & 0.57      & 0.56       & 0.60       & 0.58       & 0.56       & 0.60       & 0.54      & 0.55       & 0.54      & 0.55      & 0.55       & 0.56      & 0.55      & 0.56       & 0.55      \\
          & \cpm 0.04 & \cpm  0.04 & \cpm  0.04 & \cpm 0.05 & \cpm 0.05  & \cpm 0.05  & \cpm 0.05  & \cpm 0.05  & \cpm 0.05  & \cpm 0.04 & \cpm 0.04  & \cpm 0.04 & \cpm 0.04 & \cpm 0.04  & \cpm 0.05 & \cpm 0.03 & \cpm 0.04  & \cpm 0.04 \\
2         & 0.98      & 0.98       & 0.98       & 0.96      & 0.95       & 0.98       & 0.96       & 0.95       & 0.98       & 0.98      & 0.99       & 0.98      & 0.99      & 0.99       & 1.01      & 1.01      & 1.03       & 1.00      \\
          & \cpm 0.06 & \cpm 0.06  & \cpm 0.06  & \cpm 0.06 & \cpm 0.06  & \cpm 0.06  & \cpm 0.06  & \cpm 0.05  & \cpm 0.06  & \cpm 0.06 & \cpm 0.06  & \cpm 0.06 & \cpm 0.06 & \cpm 0.06  & \cpm 0.06 & \cpm 0.05 & \cpm 0.06  & \cpm 0.06 \\
3         & 1.25      & 1.26       & 1.24       & 1.22      & 1.22       & 1.21       & 1.21       & 1.21       & 1.21       & 1.31      & 1.34       & 1.34      & 1.31      & 1.34       & 1.38      & 1.36      & 1.41       & 1.38      \\
          & \cpm 0.06 & \cpm 0.06  & \cpm 0.06  & \cpm 0.04 & \cpm 0.03  & \cpm 0.04  & \cpm 0.03  & \cpm 0.03  & \cpm 0.04  & \cpm 0.08 & \cpm 0.08  & \cpm 0.07 & \cpm 0.07 & \cpm 0.06  & \cpm 0.06 & \cpm 0.06 & \cpm 0.06  & \cpm 0.07 \\
4         & 1.43      & 1.44       & 1.38       & 1.38      & 1.392      & 1.34       & 1.364      & 1.382      & 1.33       & 1.54      & 1.58       & 1.63      & 1.55      & 1.61       & 1.69      & 1.62      & 1.70       & 1.71      \\
          & \cpm 0.04 & \cpm 0.04  & \cpm 0.06  & \cpm 0.01 & \cpm 0.009 & \cpm 0.03  & \cpm 0.005 & \cpm 0.004 & \cpm 0.02  & \cpm 0.09 & \cpm 0.09  & \cpm 0.08 & \cpm 0.07 & \cpm 0.05  & \cpm 0.07 & \cpm 0.07 & \cpm 0.06  & \cpm 0.09 \\
5         & 1.53      & 1.55       & 1.45       & 1.46      & 1.49       & 1.384      & 1.44       & 1.47       & 1.370      & 1.7       & 1.8        & 1.9       & 1.71      & 1.81       & 1.95      & 1.8       & 1.9        & 2.0       \\
          & \cpm 0.03 & \cpm 0.03  & \cpm 0.05  & \cpm 0.01 & \cpm 0.01  & \cpm 0.008 & \cpm 0.03  & \cpm 0.03  & \cpm 0.007 & \cpm 0.1  & \cpm 0.1   & \cpm 0.1  & \cpm 0.06 & \cpm 0.08  & \cpm 0.08 & \cpm 0.1  & \cpm 0.1   & \cpm 0.1  \\
\hline \hline
\end{tabular*}
\end{table*}

Figure~\ref{fig:width_mp} shows the absolute width $\tau_2 - \tau_1$ of the estimated inertial range (using the criteria described above) as a function of the number of times $n$ used in the multi-time second-order structure function. For both VK and TG flows, and for all Cartesian components of the velocity considered, increasing the number of times employed in the structure function enlarges the width of the inertial range with respect to that obtained from $\Smp{2}{2}$. The increase from $n=2$ to $n=3$ is significant, and in most cases it doubles the width of the range of time scales compatible with inertial-range scaling, while further increasing the number of times from $n=3$ to $n=4$ results in a smaller improvement in some cases, and in saturation in others. These improvements ultimately result in a better determination of inertial-range scaling and of the scaling exponents, as can be seen from comparing the estimated uncertainty for each scaling
exponent in both flows considered, and for different values of $n$, shown in table \ref{table2}.

\subsection{Reynolds number dependence}

\begin{figure}[t!]
{\includegraphics[width=.85\textwidth]{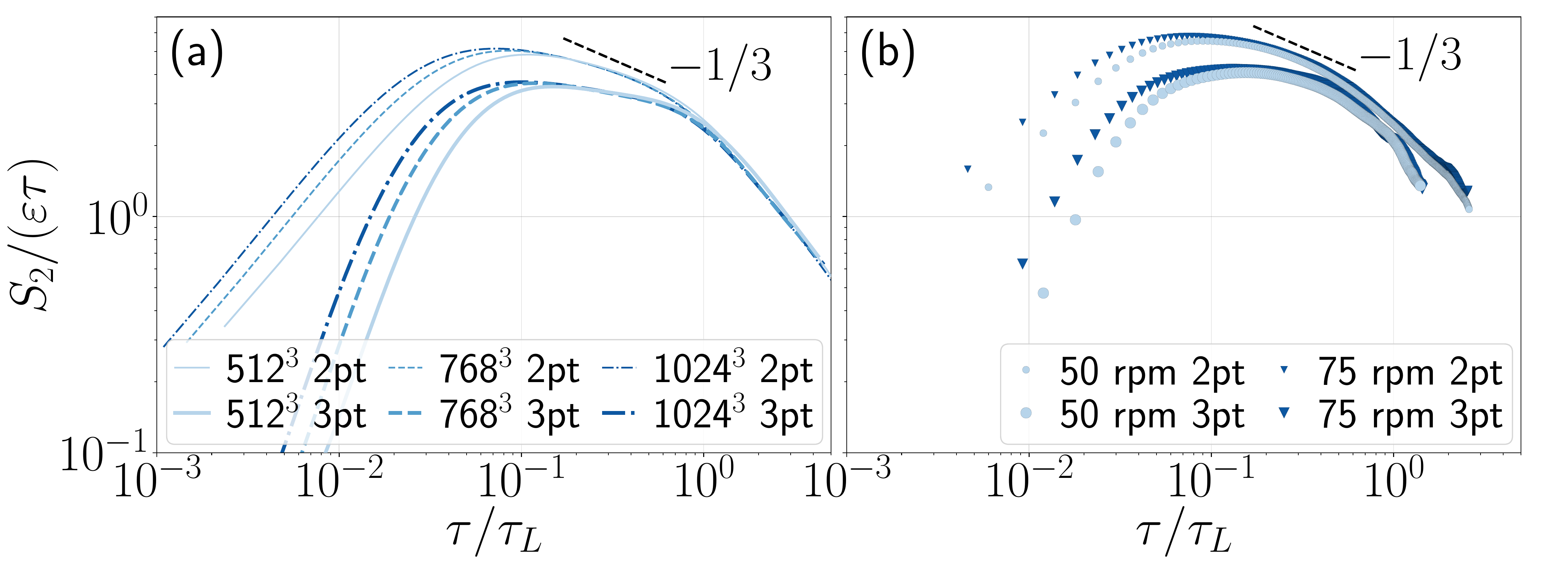}}%
\hfill
\caption{Second-order structure functions using 2 and 3 times compensated by the inertial range prediction, for all values of the Reynolds number considered in this study. Panel (a) shows the structure function for the TG simulations, labeled according to grid resolution. Panel (b) shows the data corresponding to the VK experiment, for different rotation frequencies of the propellers. Larger number of grid points in DNSs and higher rotation frequencies in experiments correspond to higher Reynolds numbers (see table \ref{table1}). A power law with exponent $-1/3$ -which corresponds to sweeping by the mean flow given by $S_2 \propto \tau^{2/3}$- is indicated by the dashed line.}
\label{fig:Re_dep}
\end{figure}

So far we considered only the VK experiment and the TG DNS at the largest available Reynolds number for each case. In the two types of swirling flows considered, the effect of the mean flow on the Lagrangian statistics should be similar for different values of this number, since the macroscopic structures that make up the mean flow depend mostly on the energy injection mechanism and the boundary conditions, which remain the same as $\textrm{Re}$ is increased, and as a result, the mean flow should depend only weakly on $\textrm{Re}$ in the fully developed turbulent state. However, the sub-leading inertial range scaling of the Lagrangian structure functions should depend on $\textrm{Re}$. In this section we thus compare the multi-time second-order structure function for the different Reynolds numbers available in the simulations and in the experiments.

To illustrate the effect of varying the Reynolds number we show only the multi-time second-order structure function of the $x$-velocity component using 2 and 3 times; our results exhibit a similar behavior for the other Cartesian velocity components, as well as when using 4 times. In Fig.~\ref{fig:Re_dep}(a), $\Smp{2}{n}(\tau)$, using $n=2$ and 3 times and compensated by $\varepsilon \tau$ is shown for the TG DNSs with $512^3$, $768^3$, and $1024^3$ grid points, while in panel (b) of Fig.~\ref{fig:Re_dep} the curves correspond to the VK experiment with forcing frequencies $f_0^\text{VK}=50$ rpm and 75 rpm.  For fixed $n$, the overall shape of the structure functions is similar at the largest time lags for all Reynolds numbers, in agreement with the independence of the sweeping-dominated range with $\textrm{Re}$. For $n=2$, a scaling $\Smp{2}{2}(\tau) \propto \tau^{2/3}$, shown by the dashed line in Fig.~\ref{fig:Re_dep}, is apparent at those time scales, and increasing $\textrm{Re}$ also results in a small increase of the inertial range compatible with Lagrangian Kolmogorov scaling, $S_2 (\tau) \sim \varepsilon \tau$, for shorter times.

Increasing the number of times used in the computation of the multi-time structure functions reduces the effect of the large-scale flow in the scaling, in a similar way for all $\textrm{Re}$, and as already discussed results in a faster drop of the curves at the dissipative scales. When using $3$ times, the compensated plateau compatible with $\Smp{2}{3}(\tau) \propto \tau$ scaling increases with increasing $\textrm{Re}$, and when comparing TG and VK data at the largest available $\textrm{Re}$, is slightly wider for the VK data. This results from a slightly larger $\textrm{Re}$ in the experiments, and from the different spectral widths of the forcing mechanisms in the experiments and the simulations. These results confirm the capacity of the multi-time structure functions to reduce large-scale contamination effects on the Lagrangian statistics in flows with different forcing mechanisms, without the need of {\it a priori} knowledge of the large-scale flow geometry or the boundary conditions. Moreover, it is worth mentioning that even in idealized conditions (e.g., when using delta-correlated in time isotropic forcing, as is usual when studying HIT numerically), the effect of random sweeping by large-scale eddies is not negligible, and is know to also contaminate the inertial-range scaling at finite Reynolds numbers, at least in the case of Eulerian studies \cite{Chen_1989}.

\subsection{Velocity power spectra}

A common physical interpretation of the two-times second-order structure function is associated with its relation to the energy spectrum through the Wiener-Khinchin theorem. As a result, $S_2(\tau)$ (or $S_2(\ell)$ in the Eulerian case) is associated, sometimes misleadingly, to the energy or the variance of the velocity field at a given scale. As discussed here and thoroughly in the literature (see, e.g., \cite{Davidson_2015}), structure functions at a given scale also have contributions from slower (or larger) eddies. Structure functions are also not only relevant to study intermittency, but as a way to estimate the energy spectrum from experiments, for which data is not periodic in space or in time. Thus, the question of how the multi-time structure functions are related with the energy spectrum has some relevance. In particular considering that in spectral space, the inertial-range scaling has been reported to be under some circumstances less affected by sweeping. As an example, in laboratory experiments a $-2$ power-law scaling (compatible with a scaling $\propto \tau$ of the structure function) has been reported for the Lagrangian velocity power spectrum \cite{Mordant_2004}.

We can expand the 3-times second-order structure function 
$\Smp{2}{3}(\tau) = \frac{1}{3} \langle [v(t+\tau) - 2v(t) + v(t-\tau)]^2\rangle$ 
and rewrite it in terms of the velocity correlation function $C_v(\tau)$, to obtain
\begin{equation}
	\Smp{2}{3}(\tau) = 2~C_v(0) - \tfrac{8}{3}~C_v(\tau) + \tfrac{2}{3}~C_v(2\tau).
	\label{eq:S3_Cv_link}
\end{equation}
Note that this expression is the 3-times extension of the standard relation $S_2(\tau) = 2 \, C_v(0) - 2 \, C_v(\tau)$, linking the velocity correlation function to the 2-times second-order structure function. By isolating the second term on the right hand side of equation~\eqref{eq:S3_Cv_link},
\begin{equation}
	C_v(\tau) = \tfrac{3}{4}~C_v(0) - \tfrac{3}{8}~\Smp{2}{3}(\tau) + \tfrac{1}{4}~C_v(2\tau),
\end{equation}
it is straightforward to obtain an expression for the energy spectrum by means of the Wiener-Khinchin theorem. As this spectrum is based on $\Smp{2}{3}(\tau)$, we symbolize it as $E^{3\text{pt}}(f)$; its definition being
\begin{equation}
	E^{3\text{pt}}(f) = \tfrac{3}{4}~C_v(0)   - 
	\tfrac{3}{8}~\widehat{ \Smp{2}{3}(\tau) } + 
	\tfrac{1}{8}~E^{2\text{pt}}(f/2),
    \label{eq:E3p}
\end{equation}
where the hat operator ($\:\widehat{\cdot}\:$) indicates the cosine transform using a Blackman window, and where $E^\text{2pt}(f)$ represents the usual power spectrum calculated as the transform of the velocity correlation function, i.e., $E^{2\text{pt}}(f) = \widehat{C_v(\tau)} \equiv C_v(0) - \widehat{\Smp{2}{2}(\tau)}/2$. In this context, the 3-times energy spectrum can be thought of as a  correction to the standard, 2-times based, energy spectrum. Following this procedure, expressions similar to Eq.~\eqref{eq:E3p} can be derived for $E^{n \text{pt}}(f)$ which would involve multi-time structure functions computed using a number of times less than or equal to $n$.

\begin{figure}[t!]
  {\includegraphics[width=0.5\textwidth]{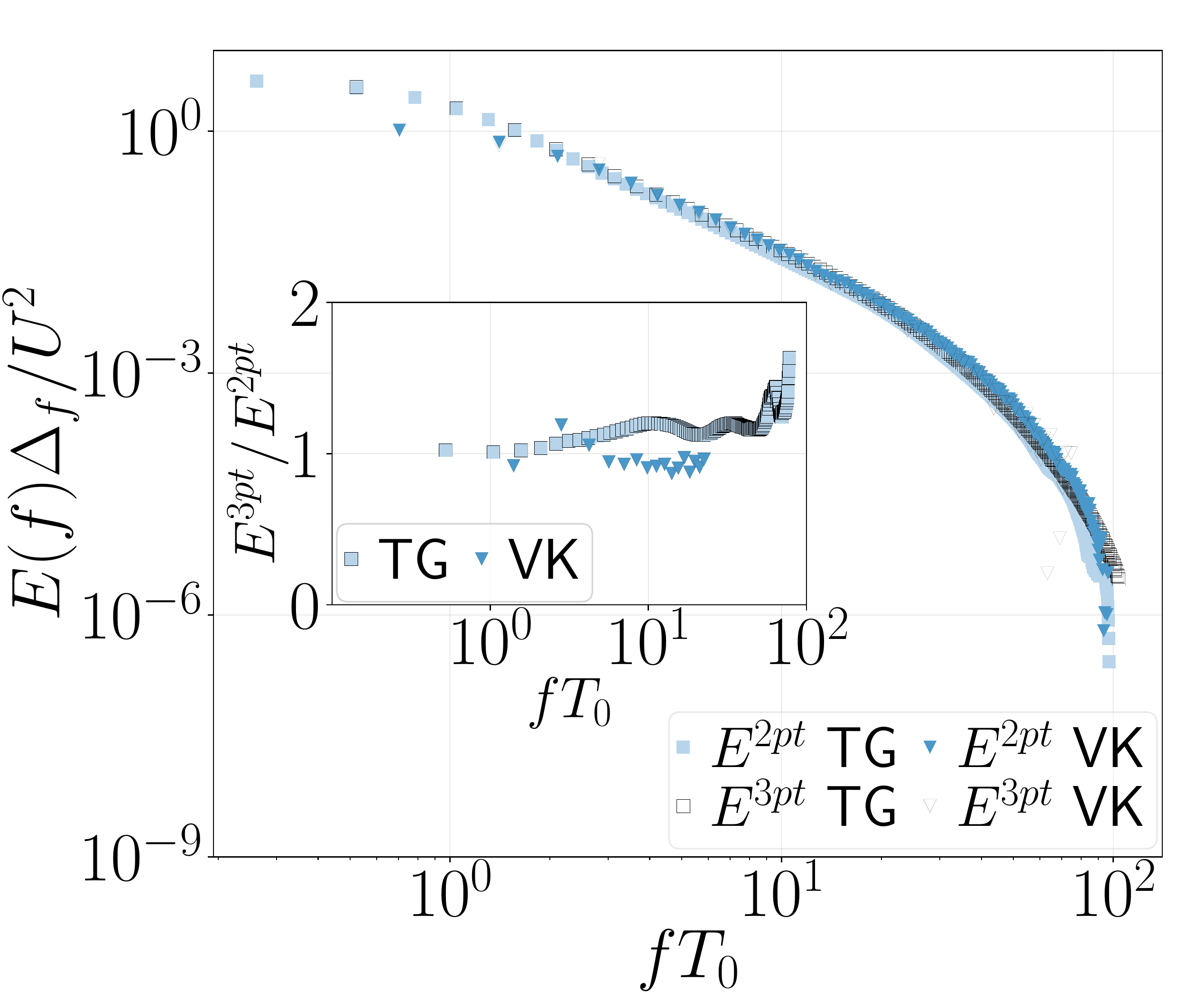}}
  \hfill
  \caption{Lagrangian velocity power spectra computed from the velocity correlation function ($E^\text{2pt}$, filled markers) and from the 3-times second-order structure function ($E^\text{3pt}$, empty markers), in log-log scale. Square and triangle markers denote results from simulations and experiments, respectively. The inset shows $E^{3\text{pt}}/E^{2\text{pt}}$ for each dataset in log-lin scale.}
  \label{fig:energy_spec}
\end{figure}

Figure~\ref{fig:energy_spec} shows $E^{3\text{pt}}(f)$ computed as in Eq.~\eqref{eq:E3p}, compared to $E^{2\text{pt}}(f)$ obtained from the 2-times second-order structure function, for an arbitrary Cartesian component of the Lagrangian velocity (similar results are observed for the other components). The spectra computed using 3 times and the one obtained from the correlation function (or equivalently, from the 2-times second-order structure function) are in good agreement. The quotient $E^{3\text{pt}}/E^{2\text{pt}}$ is shown in the inset of Fig.~\ref{fig:energy_spec}. For the numerical data, the 3-times energy spectrum has slightly more energy at intermediate frequencies, which correspond to the Lagrangian inertial range. These results are in agreement with the VK data within experimental uncertainties. 

Besides showing that the spectrum can be reconstructed from the multi-time structure functions, the expressions derived here illustrate explicitly that both the 2-times structure functions as well as the energy spectrum mix with different weights contributions from different scales.

\section{Discussion}

In this work we studied multi-time structure functions (or high-order differences) of the Lagrangian velocity, and their moments, considering particle data from two turbulent flows: a von K\'arm\'an experiment, and Taylor-Green direct numerical simulations. In previous studies it was observed that the poor agreement between theory and data in the scaling of the Lagrangian inertial range \cite{Mordant_2001, Biferale_2005, Sawford_2011, Falkovich_2012} may be the result of large-scale flow effects \cite{Mordant_2002, Blum_2010, Blum_2011, Wilczek_2013}. In a recent study, we pointed out how the effect of sweeping by the macroscopic swirling flow (common to both flows considered in the present study) might be responsible for the anomalous scaling observed in the Lagrangian second-order structure function \cite{Angriman_2020}. In particular, we found a scaling proportional to $\tau^{2/3}$ for time lags $\tau \lesssim \tau_L$, consistent with sweeping of fluid elements by the large-scale eddies, and indicative of the fact that sweeping plays a relevant role in the particles' evolution even in the Lagrangian frame. Motivated by these results, in this study we considered the multi-time Lagrangian structure functions as a way to quantify the effect of sweeping on Lagrangian statistics, and to disentangle its contribution from inertial-range scaling, further extending the analysis to also consider Lagrangian intermittency. In the Eulerian case, multi-point structure functions were already shown to reduce the effect of sweeping and to provide better results for the determination of steep scaling laws (see, e.g., \cite{Falcon_2010, Chevillard_2012, Cho_2019}).

To this end, we first considered the multi-time structure functions of order $2$, which we computed for 3 and 4 times, and compared to the standard 2-times $\Smp{2}{2}$. Increasing the number of times leads to broader and better-delimited scaling laws, from which we found a scaling proportional to $\tau$ for time-lags smaller or of the order of the Lagrangian correlation time. It is worth noting that this scaling, compatible with the Lagrangian K41 prediction, was not present in the (standard) 2-times second-order structure function, except for a very narrow range at significantly smaller time scales. Our results show that the range of time scales for which the multi-time structure functions display the most significant differences with $\Smp{2}{2}$ corresponds to the
time-scale at which the effects of sweeping by the large-scale flow become dominant over the turbulent inertial-range contributions. This allowed us to derive a phenomenological expression for the scaling of $\Smp{2}{2}(\tau)$, and to identify a time scale at which the two effects become comparable.

The scaling of higher-order Lagrangian structure functions $S_p(\tau)$ (with $p>2$), pose a considerable challenge in experiments and simulations due to their heightened sensitivity to finite-Reynolds-number effects and finite width of their inertial range. Multi-time structure functions of order $p$ ranging from 1 to 5, using $2, 3$, and $4$ times, showed an increasing width of the range of time scales compatible with power-law scaling, and a reduced contamination from the mean flow. The values obtained from the scaling exponents are, within uncertainties, in good agreement with those reported in the literature for different flows \cite{Mordant_2004, Xu_2006, Benzi_2010, Sawford_2015}, which are often obtained using extended self similarity as a way to compensate for the narrow scaling range and the poor statistics. The agreement of the scaling exponents independently of the number of times considered indicates that the large-scale flow is not the direct cause behind departures from self-similarity in Lagrangian scaling. Intermittency is an inherent property of Lagrangian turbulence, which may still result from some other subtle coupling between slow and fast time scales, or may be associated, e.g., with the time irreversibility reported in tracers' trajectories in turbulent flows \cite{Xu_2014}. Besides this, the multi-time structure functions provide for a direct and flow-agnostic method to improve scaling without the need of extended self similarity, and results in an effective broadening of the inertial range, leading to a more precise determination of the associated exponents. 

A study of the behavior of multi-time statistics on the
Reynolds number, using all our datasets spanning microscale Reynolds number in the range $\textrm{Re}_\lambda \in [215, 530]$, confirmed that the sweeping range is independent of $\textrm{Re}$, a result in agreement with a phenomenological argument and with the fact that the intensity of the large-scale flow is controlled mostly by the boundary conditions and the energy injection mechanism (even though the mean flow geometry is expected to depend weakly on $\textrm{Re}$). For structure functions using 3 and 4 times, an increase in the Reynolds number results in a widening of the inertial range, as expected. For all values of $\textrm{Re}$ considered, the multi-time structure functions also considerably decrease the contamination of the inertial scaling.

Can we expect scaling to improve indefinitely as more times are used in computation of the structure functions? Our results indicate that if a given structure function displays inertial-range scaling, increasing the number of times $n$ improves the scaling. But although it may seem advantageous to then use a larger number of times, in practice datasets have finite length in time, both in simulations and in experiments. As $n$ is gradually increased, the range of available time lags shrinks drastically. This can also result in noisier structure functions and scaling exponents due to the lack of statistical convergence associated to the reduced number of data employed in their determination. As a consequence, there is an optimal value for the number of times that should be considered, which depends on the amount of data available.

As a tool, multi-time Lagrangian structure functions succeed in reducing the sweeping effect of the large-scale flow on the Lagrangian statistics of tracers. The technique is also suitable for other types of flow geometries, as {\it a priori} knowledge of the forcing mechanisms or boundary conditions is not required for its application. These structure functions also have a relation with the energy spectrum, providing an interesting link with other tools used in the characterization of turbulence.

\begin{acknowledgments}
The authors acknowledge financial support from UBACYT Grant No.~20020170100508BA and PICT Grant No.~2018-4298.
\end{acknowledgments}

\appendix
\begin{figure}[t!]
  {\includegraphics[width=1\textwidth]{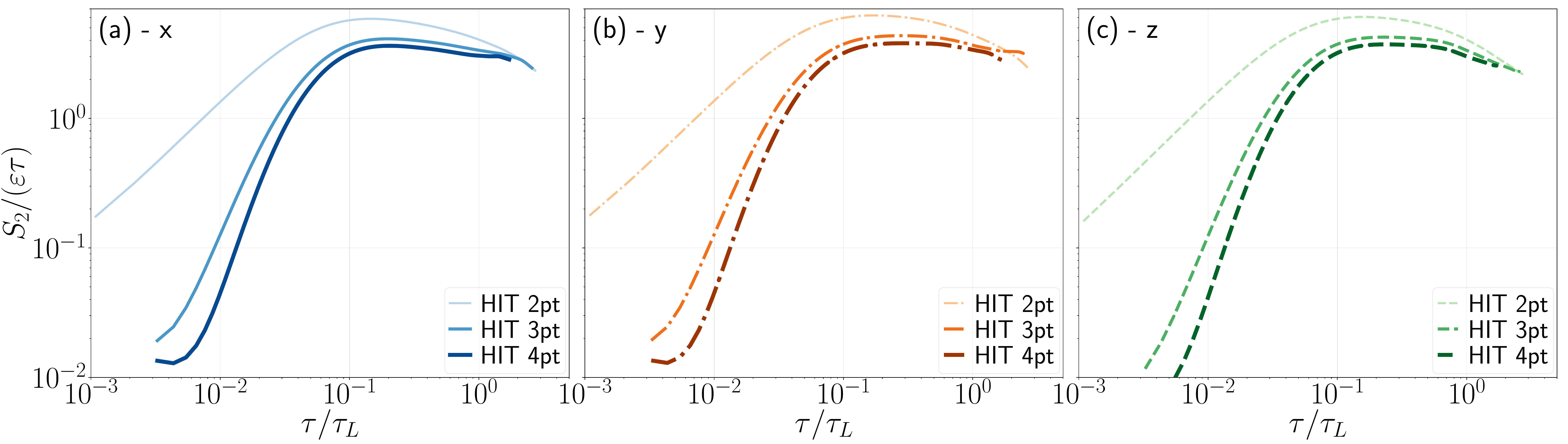}}
  \caption{Compensated second-order tracers' velocity structure functions in a $1024^3$ simulation of homogeneous and isotropic turbulence (HIT) with random forcing, using 2, 3, and 4 times, for each of the Cartesian components $x$, $y$, and $z$ of the velocity in panels (a), (b), and (c) respectively.}
  \label{fig:S2_tau_HIT}
\end{figure}

\section{Multi-time structure functions of homogeneous and isotropic turbulence}\label{app:HIT}

To analyze the performance of the multi-time structure functions in flows which do not present a mean large-scale flow, but do present large-scale velocity fluctuations, we performed a DNS of HIT using the same numerical methods as the ones used for the TG simulations. The incompressible Navier-Stokes equation was solved in a periodic cubic domain using $1024^3$ grid points, and turbulence was sustained by injecting energy in all modes in the vicinity of the Fourier shell with $k_F \approx 1$, with each mode having random phases with a correlation time of $0.5$ large eddy turnover times. The resulting Taylor microscale Reynolds number was $\textrm{Re}_\lambda \approx 330$. Once a turbulent steady state was reached, $10^6$ Lagrangian tracers were evolved in the flow following the same procedures used for TG DNSs. From the Lagrangian data, the second-order multi-time structure functions were computed using 2, 3 and 4 times, for each of the Cartesian components of the tracers' velocities. The compensated structure functions are shown in Fig.~\ref{fig:S2_tau_HIT}. An improvement in the scaling is observed when using 3 and 4 times, indicating multi-time statistics also reduce large-scale flow contamination in HIT. Similar results were obtained for higher-order structure functions.

Note that the scaling in Fig.~\ref{fig:S2_tau_HIT} improves gradually as the number of times used in the analysis is increased. The differences in the inertial range between $\Smp{2}{2}$ and $\Smp{2}{3}$ are significant. In comparison, differences between $\Smp{2}{3}$ and $\Smp{2}{4}$ are smaller, but still noticeable. The reason for this is that using $n$ times does not remove all slow (or smooth) flow components, but only those that can be described by polynomials of order up to $n$-2. Thus, improvements (albeit gradually smaller) can still be expected as the number of times is further increased. Interestingly, $\Smp{2}{3}$ and $\Smp{2}{4}$ seem to converge faster (i.e., to be less different) in the inertial ranges of the TG flow and of the VK experiment than in the case of HIT. This can be expected, as the randomness of the forcing used in HIT, when compared with the mean, smooth, and large-scale flow in the TG flow and VK experiment, can require higher-order polynomial contributions to successfully remove the sweeping. As a result, the differences between $\Smp{2}{3}$ and $\Smp{2}{4}$ in the inertial range in Fig.~\ref{fig:S2_tau_HIT} indicate that some contamination due to the sweeping remains. In spite of this, the multi-time structure functions result in a more clear inertial range scaling when compared with $\Smp{2}{2}$.

\section{Multi-time scaling exponents using Extended Self-Similarity}\label{app:ESS}

As in the literature the scaling exponents of Lagrangian structure functions are usually reported using Extend Self-Similarity (ESS), for an easier comparison we also performed the calculation of the exponents of the multi-time structure functions using ESS, $\lse{p}{ESS, n}$, using $2$, $3$ and, $4$ times. The exponents were calculated by performing a fit of $\Smp{p}{n}$ as a function of $\Smp{2}{n}$ (with $p=1,\,3,\,4$, and $5$). Their 95\% confidence intervals were estimated from the fit. The exponents and their uncertainties are listed in table~\ref{table_ESS}. The values are compatible with the ones previously reported using ESS in the literature, and with the exponents computed using Eq.~(\ref{eq:LSE}) and shown in table~\ref{table2}, but with smaller uncertainties, as expected when assuming ESS.

\begin{table*}[t!]
\caption{Scaling exponents (LSEs) computed using ESS, $\xi_p^{\text{ESS}}$, of orders $p=1,\,3,\,4$, and $5$ for the $x$, $y$, and $z$ components of the Lagrangian velocity in the TG and VK datasets, using 2, 3, and 4 times. Uncertainties are listed below each exponent value.}
\label{table_ESS}
\renewcommand*{\arraystretch}{1.4}		
\newcommand*{\cpm}{{\scriptstyle\pm}}	
\scriptsize
\begin{tabular*}{\textwidth}{@{\extracolsep{\fill}} c *{9}{d{1.3}} *{9}{d{1.3}} }
\hline \hline \hfill
		  &           &            &            &           & \mc{TG}    &            &            &            &            &           &            &           &           & \mc{VK}    &           &           &            &           \\
\cmidrule[0.4pt]{2-10} \cmidrule[0.4pt]{11-19} 
		  &           & \mc{2pt} &            &           & \mc{3pt} &            &            & \mc{4pt} &            &           & \mc{2pt} &           &           & \mc{3pt} &           &           & \mc{4pt} &           \\
\cmidrule[0.4pt]{2-4} \cmidrule[0.4pt]{5-7} \cmidrule[0.4pt]{8-10} \cmidrule[0.4pt]{11-13} \cmidrule[0.4pt]{14-16} \cmidrule[0.4pt]{17-19}
\mc{$p$}  & \mc{$x$}  & \mc{$y$}   & \mc{$z$}   & \mc{$x$}  & \mc{$y$}   & \mc{$z$}   & \mc{$x$}   & \mc{$y$}   & \mc{$z$}   & \mc{$x$}  & \mc{$y$}   & \mc{$z$}  & \mc{$x$}  & \mc{$y$}   & \mc{$z$}  & \mc{$x$}  & \mc{$y$}   & \mc{$z$}  \\
\cmidrule[0.4pt]{1-19}
1         &    0.5821 &     0.5783 &     0.5903 &     0.600 &     0.589 &     0.614 &     0.603 &     0.592 &     0.614 &     0.5563 &     0.5539 &     0.554 &     0.559 &     0.551 &     0.555 &     0.549 &     0.543 &     0.549 \\
$\pm$          & 0.0008 &  0.0009 &  0.0005 &  0.002 &  0.002 &  0.002 &  0.002 &  0.002 &  0.002 &  0.0008 &  0.0007 &  0.001 &  0.002 &  0.002 &  0.003 &  0.001 &  0.002 &  0.003 \\
3         &     1.277 &      1.290 &      1.258 &     1.259 &     1.287 &     1.237 &     1.254 &     1.284 &     1.236 &     1.3375 &      1.344 &     1.361 &     1.332 &     1.356 &     1.369 &     1.346 &     1.367 &     1.382 \\
$\pm$          &  0.002 &   0.003 &   0.002 &  0.003 &  0.003 &  0.003 &  0.003 &  0.003 &  0.004 &  0.0009 &   0.001 &  0.003 &  0.002 &  0.005 &  0.006 &  0.002 &  0.006 &  0.003 \\
4         &     1.449 &      1.485 &      1.407 &     1.417 &     1.489 &     1.374 &     1.405 &     1.482 &     1.370 &      1.578 &      1.594 &     1.659 &     1.567 &      1.63 &      1.68 &     1.595 &      1.65 &     1.714 \\
$\pm$          &  0.005 &   0.006 &   0.005 &  0.006 &  0.007 &  0.007 &  0.007 &  0.007 &  0.007 &   0.001 &   0.002 &  0.005 &  0.005 &   0.01 &   0.01 &  0.008 &   0.02 &  0.006 \\
5         &     1.547 &       1.62 &      1.485 &     1.502 &      1.63 &      1.44 &      1.48 &      1.62 &      1.43 &      1.730 &      1.761 &     1.919 &     1.718 &      1.84 &      1.94 &      1.76 &      1.86 &      1.99 \\
$\pm$          &  0.008 &    0.01 &   0.007 &  0.009 &   0.01 &   0.01 &   0.01 &   0.01 &   0.01 &   0.002 &   0.002 &  0.005 &  0.009 &   0.03 &   0.02 &   0.02 &   0.04 &   0.01 \\

\hline \hline
\end{tabular*}
\end{table*}

\bibliography{ms}
\end{document}